# Spectroscopic and first principle DFT+eDMFT study of complex structural, electronic, and vibrational properties of $M_2Mo_3O_8$ ($M$=Fe, Mn) polar magnets

T. N. Stanislavchuk,[1,a)] G. L. Pascut,[2,3] A. P. Litvinchuk,[4] Z. Liu,[5] Sungkyun Choi,[2] M. J. Gutmann,[6] B. Gao,[2] K. Haule,[2] V. Kiryukhin,[2,7] S.-W. Cheong,[7] and A. A. Sirenko[1]

[1] *Department of Physics, New Jersey Institute of Technology, Newark, New Jersey 07102, USA*

[2] *Department of Physics and Astronomy, Rutgers University, Piscataway, New Jersey 08854, USA*

[3] *MANSiD Research Center and Faculty of Forestry, Stefan Cel Mare University (USV), Suceava 720229, Romania*

[4] *Texas Center for Superconductivity and Department of Physics, University of Houston, Houston, Texas 77204, USA*

[5] *George Washington University, Washington, District Columbia 20052, USA*

[6] *ISIS Facility, STFC-Rutherford Appleton Laboratory, Didcot OX11 OQX, United Kingdom*

[7] *Rutgers Center for Emergent Materials and Department of Physics and Astronomy, Rutgers University, Piscataway, New Jersey 08854, USA*

---

a) Author to whom correspondence should be addressed. Electronic mail: stantar@njit.edu.

ABSTRACT

Optical spectroscopy, X-ray diffraction measurements, density functional theory (DFT) and density functional theory + embedded dynamical mean field theory (DFT+eDMFT) have been used to characterize structural and electronic properties of hexagonal $M_2Mo_3O_8$ ($M$ = Fe, Mn) polar magnets. Our experimental data are consistent with the room temperature structure belonging to the space group $P6_3mc$ for both compounds. The experimental structural and electronic properties at room temperature, are well reproduced within DFT+eDMFT method, thus establishing its predictive power in the paramagnetic phase. With decreasing temperature, both compounds undergo a magnetic phase transition and we argue that this transition is concurrent with a structural phase transition (symmetry change from $P6_3mc$) in the Fe compound and an isostructural transition (no symmetry change from $P6_3mc$) in the Mn compound. In addition, the unusual temperature dependent behavior of electronic $d$-$d$ transitions in $Fe^{2+}$ ions is discussed.

# I. INTRODUCTION

Materials with coupled magnetic and electric degrees of freedom have attracted significant attention due to their importance in designing of novel electronic devices, such as magnetoelectric (ME) tunneling junctions, weak magnetic field sensors, microwave phase shifters, low-energy consuming electronics and many other applications [1,2,3,4]. For ME effect to be observed, the time reversal and space inversion symmetries should be broken. This is always fulfilled in multiferroics, i.e. materials with simultaneous magnetic and electric orders, which stimulated a growing interest to their properties. Many multiferroics are multi-domain compounds, which weakens the total ME effect averaged among all domains. Special poling procedures are required to achieve their full ME potential. In contrast, polar magnets, i.e. materials which crystallize in a polar structure and possess magnetic order, can often be grown as mono-domain, which is beneficial for ME applications. In this paper, using experimental and theoretical techniques, we study the representative compounds of the $M_2Mo_3O_8$ ($M$ is a transition metal) polar magnet family [5, 6], which possess strong spin-lattice coupling resulting in rich ME properties. In particular, a large tunable ME effect was reported for $M_2Mo_3O_8$ ($M$=Fe, Mn) polar magnets in both static [7,8,9] and dynamic regimes [10,11,12]. Also, a giant thermal Hall effect has recently been observed in $Zn_xFe_{1-x}Mo_3O_8$ crystals revealing the influence of spin-lattice coupling on low-energy acoustic phonon modes. [13].

ME properties of the system can be described with a ME tensor, the symmetry of which is determined by crystallographic and magnetic symmetries. We have found that at room temperature the crystal structure of the $M_2Mo_3O_8$ belongs to a polar $P6_3mc$ space group and consists of $M^{2+}$ and $Mo^{4+}$ layers stacked along the $c$-axis [see Fig. 1(a)]. The $M^{2+}$ layer is composed of the corner-sharing $MO_6$ octahedra and $MO_4$ tetrahedra, the orientation of the vertices of the latter determines the direction of electric polarization along the $c$-axis. Magnetic $M^{2+}$ layers are separated by non-magnetic trimerized $MoO_6$ octahedra [see Fig. 1(b)]. Both compounds, $Mn_2Mo_3O_8$ and $Fe_2Mo_3O_8$, order magnetically below $T_C$ = 41 K for the Mn- and $T_N$ = 60 K for the Fe- compound with the magnetic moments collinear with the $c$-axis [14]. Neutron scattering experiments showed that the magnetic structures are of antiferromagnetic and ferrimagnetic types for the Fe- and Mn- compounds, respectively, with different magnetic moments on the octahedral and tetrahedral sites [see Fig. 1(c)] [14]. Within each $M^{2+}$ layer the magnetic moments on the octahedral and tetrahedral sites are aligned antiferromagnetically, thus giving rise to a net ferromagnetic intra-layer moment [see Fig. 1(c)]. For the Mn-compound the ferromagnetic intra-layer moments are coupled ferromagnetically resulting in the ferrimagnetic (FRM) order type while for the Fe-compound these intra-layer moments are coupled antiferromagnetically, thus hiding the ferromagnetic intra-layer moment and giving rise to antiferromagnetic (AFM) order type [see Fig. 1(c)]. Recently it was shown that the ferromagnetic intra-layer moment can be revealed in the Fe-compound either by application of a magnetic field $H||c$ or by Zn doping, thus enabling switching between AFM and FRM order types, which has a substantial implication to the large magnetic-field-tunable ME effect reported in the Fe-compound [7,8,15].

Combined studies of specific heat, pyroelectric current and dielectric susceptibility suggested that ordering of $M$ spins is concurrent with structural changes in both Fe- and Mn- compounds, see Fig. 1(d) and Refs [7,9]. Also, it has been shown that the changes in polarization, $\Delta P(T)$, are similar in both Mn- and Fe- compounds [see Fig. 1(d)] [9] which implies that the atomic displacements are of the same order of magnitude. The changes in polarization for the Fe-compound have been reproduced using a simple model based on the difference between the atomic coordinates in the ordered state (obtained by DFT+U) and in the paramagnetic state (determined by experiment) [7]. Although these calculations could reproduce the order of magnitude for $\Delta P(T)$, and suggest that the origin of these structural changes are the exchange striction effects, no information about the low temperature (LT) symmetry of these materials was obtained [7], thus an open question still exists about the type of the structural phase transition.

The goals of this work are (I) to unravel the nature of these atomic displacements occurring through the magnetic transition and learn if they are consistent with breaking the high temperature (HT) symmetry (structural phase transition) or with preserving it (isostructural phase transition), and (II) to prove the predictive powers of the eDMFT method for the structural and electronic properties in the paramagnetic state.

To answer these questions, we revised the room temperature crystal structures using single crystal X-ray diffraction, performed infrared (IR) and Raman studies of phonon modes in the temperature range 300 - 5 K, and investigated the electronic properties, such as the band gap magnitude and the crystal field levels in $M_2Mo_3O_8$ ($M$=Fe, Mn) compounds. The reason we chose spectroscopic techniques to probe the symmetry changes across the magnetic transitions is because the number of spectroscopically accessible lattice excitations is determined by the crystal's symmetry, and thus these techniques are very sensitive to phase transitions upon which the symmetry changes. To the best of our knowledge, only unpolarized IR transmission and Raman measurements on polycrystalline $M_2Mo_3O_8$ ($M$ = Co) samples have been reported in literature so far [16]. In addition, we have employed density functional theory (DFT) and density functional theory + embedded dynamical mean field theory (DFT+eDMFT) to understand better the interplay of the structural and electronic degrees of freedom in the paramagnetic state of these materials with complex crystal structures, see Fig. 1.

Based on our experimental and theoretical approaches we will show that (I) the magnetic transition is concurrent with a structural phase transition in the Fe-compound and an iso-structural phase transition in the Mn-compound; (II) there is an overall good agreement between the DFT calculated and room temperature experimental infrared and Raman phonon modes apart from some low frequency modes in the Fe-compound, (III) the electronic and structural properties at finite temperature (in the paramagnetic state) reproduced by DFT+eDMFT are in very good agreement with the experiment, thus confirming the predictive powers of the DFT+eDMFT method at finite temperature and (IV) the group of spectral lines observed in the IR and Raman spectra of $Fe_2Mo_3O_8$ in 3000 – 3500 cm$^{-1}$ range and demonstrating unusual temperature dependent behavior corresponds to electronic $d$-$d$ transitions in $Fe^{2+}$ ions.

## II. SAMPLES, EXPERIMENTAL TECHNIQUES, AND CALCULATIONS

$M_2Mo_3O_8$ ($M$ = Fe, Mn) and $FeZnMo_3O_8$ were grown using a chemical vapor transport method at the Rutgers Center for Emergent Materials [7]. Single crystals with naturally terminated faces had a typical size of 0.5×0.5×0.5 mm$^3$. Different samples with crystal faces that contained two different orientations of the *c*-axis, in the plane of the sample and perpendicular to it, have been selected for spectroscopic experiments. Far-infrared (FIR) reflectivity measurements were performed for the electric field of light directed along and perpendicular to the *c*-axis of the crystals in the 60–7000 cm$^{-1}$ spectral range using Bruker v80 interferometer with a glowbar light source and a 15x Schwarzschild objective. The spectral resolution in the far-IR experiments was 2 cm$^{-1}$. Micro-Raman experiments have been performed for the laser light excitation and collection along and perpendicular to the *c*-axis of the crystals in the backscattering configuration using a 532 nm laser, a LN$_2$-cooled CCD detector, and a single-grating Jobin Yvon spectrometer, which provided a spectral resolution of about 2 cm$^{-1}$. For both FIR reflectivity and Raman measurements, samples were placed in a LHe-flow optical cryostat, which allowed to perform measurements at temperatures down to 5 K. Ellipsometric measurements were performed at room temperature only in the near-IR to ultraviolet (UV) spectral range at 65º angle of incidence using J.A. Woollam M-2000 spectroscopic ellipsometer at the Center for Functional Nanomaterials at Brookhaven National Lab (CFN-BNL).

Single crystal diffraction data were collected on an Rigaku Oxford Diffraction SuperNova Diffractometer equipped with an Atlas CCD-detector and Cu Kα radiation at *T* = 300 K. Data collection, cell refinement, and data reduction were carried out using CrysAlisPro[17]. The JANA2006 software [18] was used for structure refinement. The absorption correction was done analytically using a multifaceted crystal model [19]. Extinction corrections were performed using an isotropic Becker & Coppens, Type 1, Gaussian [20]. Figures of structures were generated using VESTA [21]. Figures were prepared in Inkscape [22].

Lattice dynamics properties of Fe$_2$Mo$_3$O$_8$ and Mn$_2$Mo$_3$O$_8$ crystals were calculated within the Density Functional Theory using the *ab initio* norm conserving pseudopotential method as implemented in the CASTEP package [23]. Equilibrium atomic structures were obtained from the total-energy minimization method within DFT. Electron exchange and correlation interactions have been modeled within the local density approximation [23,24]. Plane wave basis set cut-off was 750 eV that allowed energy convergence within 10$^{-7}$ eV. The lattice was optimized until residual forces on atoms in their equilibrium positions did not exceed 5 meV/Å. Lattice dynamic properties of both compounds were further assessed via finite displacement method on a 2×2×1 supercell [25]. Integration over the Brillouin zone was performed over the 5×5×2 Monkhorst-Pack grid in reciprocal space [26]. For both materials no imaginary modes were predicted by the theory.

For calculating electronic properties and optimizations of internal coordinates, we used both (I) the density functional theory as implemented in Wien2k package [27] and (II) a fully charge-self-consistent dynamical mean-field theory method, as implemented in Rutgers DFT+eDMFT code [28,29,30]. Throughout the rest of the paper we will refer to DFT+eDMFT method as eDMFT. For the DFT part, we used the generalized gradient approximation Perdew-Burke-Ernzerhof (GGA-PBE) functional [31], RKmax = 7.0, and 312 $k$-points in the irreducible part of the 1$^{st}$ Brillouin zone. For optimizations of internal coordinates [32,33], a force criterion of 10−4 Ry/Bohr was adopted. In order to solve the auxiliary quantum impurity problem, a continuous-time quantum Monte Carlo method in the hybridization-expansion limit (CT-HYB) was used [34], where the 5 $d$-orbitals for the Mn and Fe ions (grouped according to the local point group symmetry) were chosen as our correlated subspaces in a single-site eDMFT approximation. For the CT-HYB calculations, up to 400 million of Monte Carlo steps were employed for each Monte Carlo run. In all runs the temperature was set to be 500 K.

To define the eDMFT projector, we used the quasi-electronic orbitals by projecting bands in the large hybridization window (-10 to +10 eV with respect to the Fermi level), in which partially screened Coulomb interaction has values of $U$ = 10 eV and $J_H$ = 1 eV in both Mn and Fe ions. A nominal double counting scheme was used [35], with the $d$-orbital occupations for double counting corrections for Mn and Fe chosen to be 5 and 6, respectively.

### III. EXPERMENTAL AND THEORETICAL RESULTS

#### A. HT structural properties

The room temperature crystal structures of $Mn_2Mo_3O_8$ and $Fe_2Mo_3O_8$ compounds have been revised using single crystal X-ray diffraction. Our findings at the room temperature are consistent with the previously reported space group symmetry and structural parameters [6,36,37,38,39]. The good quality of our refinements can be seen from the plot of the observed vs. calculated structure factors squared, $|F|^2$, satisfying $|F|^2 > 3\sigma(|F|^2)$ shown in Figs. 2(a) and 2(b). From the total number (5140/4325 for $Mn_2Mo_3O_8$/$Fe_2Mo_3O_8$) of measured intensities, we obtained 271/271 unique reflections satisfying $|F|^2 > 3\sigma(|F|^2)$ for $Mn_2Mo_3O_8$/$Fe_2Mo_3O_8$ compounds that are contributing to the least-squares refinements calculation (reflections related by symmetries are merged together). The ratio data/parameters used in the "Full least-squares on $F^2$" refinement method was 271/32 for each compound. Other parameters commonly used to characterize the data refinements are shown as insets in Figs. 2(a) and 2(b). These parameters are defined as follows: (I) the residual factor, $R_1$, for the reflections used in the refinements, $R_1 = \frac{\sum |F_{obs} - F_{calc}|}{\sum |F_{obs}|}$; (II) the weighted residual factor, $wR_2$, for the reflections used in the refinements, $wR_2 = \sqrt{\frac{\sum w(F_{obs}^2 - F_{calc}^2)^2}{\sum (F_{obs}^2)^2}}$, where $w = \frac{1}{[\sigma(F_{obs}^2)]^2 + 0.0004 F_{obs}^4}$; (III) the goodness of fit $S = \sqrt{\frac{\sum w(F_{obs}^2 - F_{calc}^2)^2}{N_{ref} - N_{param}}}$. In these equations, $F_{obs}$ and $F_{calc}$ represent the observed and calculated structure factors, $N_{param}$ represents the number of refined parameters, $N_{ref}$ represents the number of unique reflections used in the refinement and the sum is taken over all $N_{ref}$ reflections.

Our crystals reveal the high-quality by showing that more than 98% of detected reflections are indexed by a single hexagonal domain corresponding to the $P6_3mc$ hexagonal symmetry. However, we also observed a few very weak forbidden reflections within the $P6_3mc$ space group (No. 186). If these reflections are genuinely observable nuclear Bragg reflections then the potential space group describing these systems should be from the trigonal symmetry such as $P3m$ (No. 156) or $P3$ (No. 143), where all the observed reflections are allowed by symmetry and can be indexed. As this hypothesis can be an important result related to the interplay between the true crystal structure and electronic responses already at room temperature, we designed the azimuthal-angle measurements to understand the nature of these forbidden reflections. We chose four representative reflections covering both allowed and forbidden reflections using the $Fe_2Mo_3O_8$ single crystal according to the refection conditions for the general Wyckoff site. The results of the azimuthal scan are presented in Fig. 2(c) where we show the evolution of normalized intensity for a wide range of azimuthal angles $\Psi_i$ ($i \in [1, n]$ where $n$ is a total number of measured angles, see Fig. 2(c)). Each intensity $I_i$ was first divided by the corresponding sigma $\sigma_i$ to get statistically better-defined parameters, and then $(I_i/\sigma_i)$ was further normalized by the average intensity $I_{av} = \frac{1}{n}\sum_{i=1}^{n} I_i$. By comparing the normalized values of intensity $(I_i/\sigma_i)/I_{av}$, as shown in Fig. 2(c), we were able to see that the normalized values for the allowed reflections are close to the unity as expected, whereas their counterpart values from forbidden reflections are strongly fluctuating upon the azimuthal angle. This is reminiscent of the observation of forbidden reflections from the unavoidable multiple diffractions, ruling out that the forbidden reflections should be considered to determine the crystal structure. Therefore, within our experimental resolution, the given hexagonal $P6_3mc$ space group (No. 186) well describes the crystal structures of $Fe_2Mo_3O_8$.

Structural relaxations of the internal atomic coordinates for fixed experimental lattice constants have been performed using non-spin polarized DFT and paramagnetic eDMFT methods. Structural relaxations have been performed using various starting artificial structures within subgroups of the experimental hexagonal space group $P6_3mc$ (No. 186) such as trigonal $P3m1$ (No. 156), $P3$ (No. 143), and even lower space groups up to monoclinic symmetry. In each case the stable relaxed crystal structures were found to have the hexagonal symmetry, space group $P6_3mc$ (No. 186) within the error bar of the calculations. While DFT is a zero-temperature method, the eDMFT is a finite-temperature method, thus eDMFT calculations were performed in the wide temperature range and the reported results in this paper correspond to the high temperature paramagnetic state ($T_{theory}$ = 500 K). In Table I, we give the relaxed internal parameters obtained experimentally and theoretically using the DFT ($T$ = 0) and eDMFT ($T$ = 500 K).

From the results presented in Table I, we see that the agreement between the experimental fractional coordinates and those obtained from the eDMFT theoretical relaxations is much better than agreement between the experimental fractional coordinates and those obtained from the DFT theoretical relaxations. eDMFT method gives smaller deviations (beyond the third digit with respect to the experiment) while the DFT method gives larger deviations (beyond the second digit with respect to the experiment).

Converting these discrepancies of internal structural parameters, to displacements in Å along the three crystallographic directions, we find a maximum discrepancy for eDMFT (DFT) to be |0.06| (|0.27|) for $Mn_2Mo_3O_8$ and |0.06| (|0.21|) for $Fe_2Mo_3O_8$. We also compute the percentage difference between the experimental and theoretically obtained values in the form $\%_{err} = \frac{1}{N}\sum \frac{|Q_{Theory} - Q_{Exp}| \cdot 100}{Q_{Exp}}$, where $Exp$ stands for experimental, $Theory$ stands for the theoretical values obtained from structural relaxations and $Q$ is any of the $x$, $y$ or $z$ fractional coordinates and the sum is taken over all $N$ internal atomic coordinates that are allowed to displace during the structural relaxations. The computed percentage error $\%_{err}$ is 0.53% (2.3%) for $Mn_2Mo_3O_8$ versus 0.32% (1.64%) for $Fe_2Mo_3O_8$ in eDMFT (DFT), correspondingly. The obtained values of discrepancies and percentage errors show that eDMFT gives much more accurate structural degrees of freedom compared to non-spin polarized DFT in these materials.

To better understand how these discrepancies between experimental and theoretical values of the fractional coordinates affect the properties of the local polyhedron formed between the central transition ion and the coordinating atoms (ligands), we have also computed a few quantities which are usually used to describe geometrically the coordination polyhedron. These quantities are the average bond length $l_{av}$(Å), polyhedral volume $V$(Å$^3$), quadratic elongation $\langle\lambda\rangle$ and bond angle variance $\sigma^2$(deg$^2$). The values of these quantities can be automatically computed using the VESTA software and besides $l_{av}$ and $V$, which have their usual meaning, the other two quantities are defined as follows: (I) quadratic elongation $\langle\lambda\rangle$ is a dimensionless quantity and gives a quantitative measure of the polyhedral distortion, independent of the effective size of the polyhedron, $\langle \lambda \rangle = \frac{1}{n}\sum_{i=1}^{n}\left(\frac{l_i}{l_0}\right)^2$, where $n$ is the coordination number of the central atom, $l_i$ is the distance from the central atom to the $i^{th}$ coordinating atom and $l_0$ is the center-to-vertex distance of a regular polyhedron of the same volume (a regular polyhedron has a quadratic elongation of 1, whereas distorted polyhedra have values greater than 1); (II) bond angle variance gives a measure of the distortion of the intra-polyhedral bond angles from the ideal polyhedron, $\sigma^2 = \frac{1}{m-1}\sum_{i=1}^{m}(\theta_i - \theta_0)^2$, where $m$ is the number of bond angles [3/2·(number of faces in the polyhedron)], $\theta_i$ is the $i^{th}$ bond angle and $\theta_0$ is the ideal bond angle for a regular polyhedron ($\theta_0$ is 90° for an octahedron and 109°28' for a tetrahedron). Bond angle variance is zero for a regular polyhedron and positive for a distorted polyhedron. The average bond length, polyhedron volume, quadratic elongation and bond angle variance are scalar quantities so they provide no information about the geometry of polyhedral distortions, but they can be used to quantitatively compare the agreement between the experimentally determined and theoretically obtained fractional coordinates.

In Figs. 3(a)-3(b) we give the values of the above-mentioned quantities for $M_2Mo_3O_8$, computed based on fractional coordinates obtained from experiment, eDMFT and DFT relaxations. In addition, we also give the percentage error between the experimental and theoretically obtained values in the form $\%_{err} = \frac{|Q_{Theory} - Q_{Exp}| \cdot 100}{Q_{Exp}}$, where $Exp$ stands for experimental, $Theory$ stands for the theoretical values obtained from structural relaxations and $Q$ can be any of the quantities, $l_{av}$, $V$, $\langle \lambda \rangle$ and $\sigma^2$. The computed percentage error ($\%_{err}$) of the quantities mentioned above, can be orders of magnitude smaller for eDMFT than DFT, see Figs 3(a)-3(b).

### B. HT electronic properties: Experiment and theory

To confirm the insulating properties of the $M_2Mo_3O_8$ compounds we have performed ellipsometric measurements of the pseudo-dielectric function in the $ab$-plane and along the $c$-axis, see Fig. 4. From our measurements we see that both compounds are insulators with across-the-gap transitions starting at 8700 cm$^{-1}$ (1.08 eV) for Fe$_2$Mo$_3$O$_8$ and 8500 cm$^{-1}$ (1.05 eV) for Mn$_2$Mo$_3$O$_8$ compounds. It was shown in the Section III.A that eDMFT method better reproduces the structural properties than DFT method does and now we will present the electronic properties obtained by these two methods. Figures 3(c)-3(d) show the total density of states obtained by the DFT method and Figs. 3(e)-3(f) show the total density of states obtained by the eDMFT method. As expected DFT gives a metallic state but eDMFT gives an insulating state with a gap ~1.2 eV for both compounds [see Figs. 3(h)-3(i)] which is close to the experimental value. If we also compute the total density of states for the relaxed crystal structures within the DFT and eDMFT methods and compare it to the density of states obtained using the experimental structure, we see that within the DFT method the two electronic structures are different but for the eDMFT method we get very similar total density of states for both compounds. The details of the electronic structures obtained by the eDMFT will be published elsewhere.

### C. LT vs HT structural properties: Infrared phonon spectra

Figure 5 shows reflectivity spectra dominated by the phonon modes for Fe$_2$Mo$_3$O$_8$ and Mn$_2$Mo$_3$O$_8$ compounds measured for two polarizations of incident light, $e\|c$ and $e\perp c$, at 85 and 5 K, i.e. above and below magnetic ordering temperatures in these compounds. To extract parameters of the phonon modes, reflectivity spectra were fitted using parametrization of the dielectric function with Drude-Lorentz oscillators [40] and the Kramers-Kronig transformation as implemented in the RefFit code [41]. In this framework, the dielectric function is parameterized as follows:

$$\varepsilon_{ab} = \varepsilon_{\infty,ab} + \sum_{i=1}^{n} \frac{\omega_{i,ab}^2 S_{i,ab}}{\omega_{i,ab}^2 - \omega^2 - i\gamma_{i,ab}\omega}$$

$$\varepsilon_c = \varepsilon_{\infty,c} + \sum_{j=1}^{m} \frac{\omega_{j,c}^2 S_{j,c}}{\omega_{j,c}^2 - \omega^2 - i\gamma_{j,c}\omega}$$

where $\omega_{i(j),ab(c)}$, $S_{i(j),ab(c)}$ and $\gamma_{i(j),ab(c)}$ are transverse optical (TO) frequency, oscillator strength and inhomogeneous broadening of the $i^{th}$ ($j^{th}$) phonon mode polarized along the *ab* plane (*c*-axis), $\varepsilon_{\infty,ab(c)}$ is the value of the dielectric function along the *ab* plane (*c*-axis) at frequencies higher than that of the highest-energy phonon mode. The extracted parameters of the phonon modes are listed in Table II. Above the magnetic ordering temperature, there are total of 19 (22) IR active phonon modes observed for $Fe_2Mo_3O_8$ ($Mn_2Mo_3O_8$) compound, 10 (9) being polarized along the *c*-axis and 9 (13) – in the *ab* plane (see Table II and spectra in Fig. 5 for $T = 85$ K). Below the magnetic ordering temperature, the number of observed phonon modes does not change for Mn compound, while three new $E_1$ modes at 270, 429 and 470 cm$^{-1}$ arise for Fe compound (see spectra in Fig. 5 for $T=5$ K). This points towards the existence of a structural phase transition, which occur concurrently with the magnetic ordering in the $Fe_3Mo_3O_8$ compound.

### D. LT vs HT structural properties: Raman phonon spectra

Spectra of the optical phonons in the same $Fe_2Mo_3O_8$ and $Mn_2Mo_3O_8$ crystals have been also studied using Raman scattering. Several back-scattering configurations were utilized: $a(c,c)\bar{a}$, $a(c,b)\bar{a}$, $c(a,b)\bar{c}$, and $c(a,a)\bar{c}$, where the first and the last symbols correspond to the *k*-vector direction for exciting and scattered light, whereas symbols in brackets correspond to the polarization direction of the exciting and scattered light. In these geometries the following mode symmetries should become accessible [42] in the $P6_3mc$ crystal structure of $M_2Mo_3O_8$ crystals: $A_1$, $E_1$, $E_2$, and $A_1+E_2$, as the in-plane *x*- and *y*-axes could not be distinguished for a hexagonal crystal. As shown in Fig. 6, the experimental Raman spectra are highly polarized, which allowed us to identify positions of the $A_1$, $E_1$ and $E_2$ phonon modes. There are total of 31 (34) Raman active modes observed for $Fe_2Mo_3O_8$ ($Mn_2Mo_3O_8$) at 85 K, among which are 9(10) $A_1$ modes, 11(12) $E_1$ modes and 11(12) $E_2$ modes. As the temperature is lowered and the Fe compound undergoes magnetic ordering, several new lines appear (see Fig. 7). In particular, two new $A_1$ modes at 232 and 852 cm$^{-1}$, one new $E_1$ mode at 747 cm$^{-1}$, and five new $E_2$ modes at 127, 158, 193, 224 and 253 cm$^{-1}$ are observed in the spectra of $Fe_2Mo_3O_8$ at 5 K. Both new $A_1$ modes at 232 and 852 cm$^{-1}$ are observed in the IR spectra at 85 K, so it is likely that they are just too weak in the Raman spectra at 85 K to be resolved. At the same time, $E_2$ Raman mode at 240 cm$^{-1}$, which is observed in the spectra at 85 K, is no longer observable at 5 K. It most likely splits into two $E_2$ modes at 224 and 253 cm$^{-1}$ at $T<T_N$. The number of modes observed for Mn compound does not change between 5 and 300 K, i.e., above and below $T_C(Mn)=41$ K. The positions and symmetries of the identified Raman active phonons for $Fe_2Mo_3O_8$ and $Mn_2Mo_3O_8$ at both 85 and 5 K are summarized in Table II along with the corresponding parameters for the IR-active modes. As expected for the polar structure of the studied compounds, there is a good agreement between positions of IR- and Raman-active phonon modes barring several modes of the $E_1$ symmetry in the $Fe_2Mo_3O_8$ compound. In the frequency range above 860 cm$^{-1}$ and up to 1300 cm$^{-1}$, we observed weak overtones for the optical phonons positioned at 1214 cm$^{-1}$ for $Fe_2Mo_3O_8$ and at 1103, 1200 and 1220 cm$^{-1}$ for $Mn_2Mo_3O_8$, which correspond to two-photon Raman scattering.

### E. HT: First principle phonon calculations

We have shown so far that eDMFT method is in a better agreement with the experimental data when it comes to the fractional atomic coordinates. In addition, eDMFT can capture the insulating state of these materials while DFT gives a metallic state. It is well known that in many cases even if DFT cannot fully explain the ground state electronic properties, the computed phonons are in good agreement with the experimentally obtained phonons. In this respect, we computed the phonon modes at the DFT level for the high temperature $P6_3mc$ structure of $M_2Mo_3O_8$ ($M$=Fe, Mn) and the obtained values for the mode frequencies are given in Table I side by side with the experimental values obtained from the IR and Raman measurements. To quantify the agreement between the calculated and measured phonon frequencies we are computing the percentage error $\%_{err} = \frac{1}{N}\sum_{i=1}^{N} \frac{|\omega^i_{\exp} - \omega^i_{theory}|}{\omega^i_{\exp}} \cdot 100$, which expresses as a percentage the difference between the computed and measured values of the phonon frequencies ($N$ is the total number of measured phonon frequencies, $\omega^i_{\exp}$ and $\omega^i_{theory}$ are the experimental and computed phonon frequencies). A percentage error very close to zero means that there is a very good agreement between the theory and the experiment. Computing the percentage error between the experimental and calculated frequencies of the IR phonon modes we obtained a percentage error of 5.3% and 3.8% for the Mn and Fe compounds, correspondingly. Another way to compare the computed and measured phonon frequencies is by calculating the mean value $\omega_{mean} = \frac{1}{M}\sum_{i=1}^{M}\omega_i$, where $M$ runs over all computed or measured phonon frequencies. The mean values for the IR phonon frequencies obtained in experiment vs. theory are 430 cm$^{-1}$ vs. 421 cm$^{-1}$ for the Mn compound and 444 cm$^{-1}$ vs. 421 cm$^{-1}$ for Fe compound. If we compute the percentage error for the mean values, we obtain 2.1% for Mn and 5.2% for Fe compounds. From the calculated values we see that the overall agreement between theory and experiment is within a few percent which we might consider as a good one.

### F. Temperature dependent electronic transitions in $Fe^{2+}$ ions in $Fe_2Mo_3O_8$

Figure 8 shows IR reflectivity spectra of $Fe_2Mo_3O_8$ in a wide spectral range from 100 to 4000 cm$^{-1}$, which covers IR active phonons located in 100-800 cm$^{-1}$ region, a broad spectral feature at ~3000 cm$^{-1}$ [Fig. 8(a)] and a group of weak narrow lines in the 3400-3500 cm$^{-1}$ region. As temperature increases from 5 to 85 K, most of the narrow lines in 3400-3500 cm$^{-1}$ region vanish, while the strong and broad feature at ~3000 cm$^{-1}$ red-shifts by several hundred wavenumbers.

Figures 9(a)-9(d) show in more detail temperature dependence of spectral lines in the 2150-3600 cm$^{-1}$ range measured for $Fe_2Mo_3O_8$ (two different samples referred to as sample_1 and sample_2 in the text below), $FeZnMo_3O_8$ and $Mn_2Mo_3O_8$ single crystals. In this spectral range the most prominent feature is the broad line at ~3000 cm$^{-1}$ observed in $Fe_2Mo_3O_8$ sample_1 [see Fig. 9(a)]. As temperature increases, the line red-shifts by ~400 cm$^{-1}$ [see inset in Fig. 9(a)] and gradually decreases in intensity but does not vanish up to room temperature. To explore the behavior of the line in more detail we measured the same spectra from a different (presumably better quality) $Fe_2Mo_3O_8$ single crystal [sample_2; see Fig. 9(b)]. Interestingly, the line was absent in the spectra of $Fe_2Mo_3O_8$ sample_2. It was also absent in the spectra of $Mn_2Mo_3O_8$ single crystal [see Fig. 9(d)] but present, although to a lesser degree, in the spectra of $FeZnMo_3O_8$ sample [see Fig. 9(c)]. The presence of this line in the spectra of $Fe_2Mo_3O_8$ and $FeZnMo_3O_8$ compounds and its absence in the spectra of $Mn_2Mo_3O_8$ compound indicates that the origin of the line is related to electronic transitions in $Fe^{2+}$ ions. As was previously reported [8] $Zn^{2+}$ ions preferably substitute $Fe^{2+}$ ions in the tetrahedral (t)

sites. Since the line is observed both in $Fe_2Mo_3O_8$ and $FeZnMo_3O_8$ samples, it should be related to the *d-d* transitions in $Fe^{2+}$ ions in octahedral (o) sites. Also, the position of this line in $FeZnMo_3O_8$ sample is blue-shifted compared to its position in $Fe_2Mo_3O_8$ sample which is probably related to the influence of the substitution of $Fe^{2+}$ ions in tetrahedral sites with $Zn^{2+}$ ions on the $Fe^{2+}$ octahedral sites.

The group of narrow lines in the 3400-3500 $cm^{-1}$ region was observed for both $Fe_2Mo_3O_8$ samples but absent in the spectra of $FeZnMo_3O_8$ and $Mn_2Mo_3O_8$ samples. Thus, we attribute these lines to *d-d* electronic transitions in tetrahedrally coordinated $Fe^{2+}$ ions.

Spectra in Fig. 9 also manifest a broad spectral feature at 3250 $cm^{-1}$ which is observed in the spectra of all studied compounds. In particular, it is clearly seen in spectra of $Fe_2Mo_3O_8$ sample_2 [Fig. 9(b)] and $Mn_2Mo_3O_8$ [Fig. 9(d)] compounds but is also observed, although hindered by overlapping with other spectral lines, in the spectra of $Fe_2Mo_3O_8$ sample_1 [Fig. 9(a)] and $FeZnMo_3O_8$ [Fig. 9(c)] samples. This feature is an instrumental artifact and is not related to the optical processes in the studied compounds.

Figure 10 shows the temperature dependence of IR and Raman spectra of $Fe^{2+}$ (t) *d-d* electronic transitions measured in different optical configurations. The positions of 5 electronic transitions have been identified and summarized in Table III for $T$ = 5 K. The lines are closely grouped with a typical distance of 10 – 20 $cm^{-1}$ from each other. With increasing temperature, intensities of the Raman-active transitions gradually decrease until they completely vanish above 52 K, i.e. near the magnetic ordering temperature $T_N(Fe) = 60$ K of $Fe_2Mo_3O_8$ [see Fig. 10(c)]. Infrared-active modes observed in Figs. 10(a)-10(b) behave in a similar way, namely they decrease in intensity with the temperature increase and are practically indistinguishable in the spectra above 70 K. At the same time as we increase the temperature two new infrared-active modes appear to the low energy side of the infrared modes observed at $T$ = 5 K and are positioned at 3358 and 3405 $cm^{-1}$ (see Fig. 11). Their intensity reaches a maximum at $T \sim 40$ K above which the lines broaden, and their intensity decreases until they disappear from the spectra above 70 K. The line at 3250 $cm^{-1}$ marked with an asterisk in Fig. 11 is due to the instrument artifact and is the same line as at 3250 $cm^{-1}$ in Figs. 9(a)-9(d).

## IV. DISCUSSION

### A. HT symmetry: Analysis of phonon modes

Room temperature single crystal X-ray diffraction together with the phonon measurements provide important information about the complex crystal structure of $M_2Mo_3O_8$ ($M$=Fe, Mn). Our single crystal X-ray diffraction measurements showed that at high temperature the crystal structure of these materials can be explained within the $P6_3mc$ space group, results which agree with the previously reported data in the literature [5,6,7].

Using the $P6_3mc$ space group, we have carried out the group theoretical analysis which predicts 21 IR active phonon modes, $9A_1 + 12E_1$, where $A_1$ modes are polarized along the $c$-axis ($e\|c$) and $E_1$ modes – perpendicular to the $c$-axis ($e\perp c$), as shown in Table IV. Our IR data measured at 85 K, i.e. at temperature where the crystal structure has the same symmetry as at room temperature, reveal 10(9) $A_1$ and 9(13) $E_1$ modes for $Fe_2Mo_3O_8$ ($Mn_2Mo_3O_8$), correspondingly, which is in good agreement with the group theory prediction, thus confirming once again the non-centrosymmetric space group $P6_3mc$ of these materials at high temperatures. Similarly, the group theory predicts 34 Raman active phonon modes, $9A_1 + 12E_1 + 13E_2$, and Raman measurements performed at 85 K reveal 9(10) $A_1$, 11(12) $E_1$ and 11(12) $E_2$ modes for $Fe_2Mo_3O_8$ ($Mn_2Mo_3O_8$) in good agreement with the prediction for the space group $P6_3mc$. The group theoretical analysis gives the number of modes and their type but does not provide information about their frequencies. To understand better our data, we performed DFT calculations of phonon modes and the computed frequencies are shown in Table II, side by side with the experimental frequencies. The crystal structure of $M_2Mo_3O_8$ crystals possessing $Fe^{2+}$ ions in both tetrahedral and octahedral coordination and $Mo^{2+}$ trimmers is rather complex and so are the underlying atomic motions corresponding to phonon modes. To have a better insight into the latter in Fig. 13 we show, as an example, the atomic displacements corresponding to two $A_1$ modes observed in the spectra of $Fe_2Mo_3O_8$. The highest frequency $A_1$ mode at 852 cm$^{-1}$ involves simultaneous deformations of both Fe- and Mo-octahedra while the most intense Raman $A_1$ mode at 446 cm$^{-1}$ is solely due to motion of $MoO_6$ octahedra.

As has been shown in the Section III.E, if we want to quantify the agreement between the theory and the experiment by the percentage error, we obtain a reasonable agreement within 6% difference. Calculating the percentage error for the individual IR phonon frequency (5.3% for the Mn and 3.8% for the Fe compound) and for the mean (2.1% for the Mn and 5.2% for the Fe compound) we conclude that the agreement is better for the Mn than for the Fe compound. Indeed, if we have a look at the phonon frequencies in Table I, we see that while for the Mn compound the frequency assignment seems to be good across the whole spectral range, for the Fe compound the calculated low energy phonon frequencies are shifted to higher energies compared to the experimental data. The reason for this discrepancy can be the fact that the ground state electronic structure is not properly accounted within non-spin polarized DFT. Since eDMFT correctly describes the insulating ground state and reproduces well the experimental structural properties it would be interesting to compute the phonon modes at the eDMFT level.

## B. HT electronic and structural predictions

Insulating materials with simultaneous magnetic and electric order, called multiferroic materials, usually belong to the class of correlated materials. Modeling the magnetic and electric properties of multiferroic materials is still a challenging problem. For example, the first step in modeling the electric polarization in these materials requires knowledge of two experimental or two theoretical crystal structures, the first one is the high temperature reference crystal structure where usually polarization is zero and the material is found in the paramagnetic state and second one is the low temperature crystal structure of the multiferroic phase where usually the atomic displacements (that give rise to finite polarization changes) with respect to the reference state are induced by long-range magnetic order. The chances of finding theoretically the two crystal structures, especially the high temperature reference structure, has improved due to the recent development of forces for correlated materials in eDMFT [43]. While the low temperature crystal structures can be obtained by spin polarized DFT, in many cases non-spin polarized DFT fails to give acceptable results for the high temperature reference structure, see for example Table I and Fig. 3. When that happens one can find an artificial spin-polarized state which could sometimes give good agreement with the experimental crystal structure even though in this case the electronic properties of the experimental paramagnetic state are misrepresented by the artificial spin-polarized state in DFT [44]. The materials studied here are pyroelectrics, which means that the structure has already built-in an electric moment due to the crystal structure and any structural changes induced by temperature, pressure, magnetic order can induce changes in the electric polarization. Thus, trying to obtain the high temperature reference crystal structure in pyroelectric materials by using an artificial spin-polarized state in DFT poses a problem since any type of magnetic order has exchange striction, where by exchange striction we mean movements of the ligand ions in order to maximize the magnetic energy gain. For example, for $Fe_2Mo_3O_8$, it has been shown in Ref. [7] that the obtained distortion pattern depends on the type of the spin-polarized DFT. Thus, it is difficult to obtain accurate fractional atomic coordinates using DFT for the pyroelectric materials.

Here we performed structural relaxations of the fractional atomic coordinates for fixed lattice parameters using DFT and eDMFT at high temperatures. As discussed in the Section III and shown in Table I and Fig. 3, eDMFT gives much better agreement with the experimental results than DFT does. In addition to structural properties, eDMFT also reproduces very well the electronic properties. For example, we see from Figs. 3(e) and 3(f) that the total density of states computed for the crystal structure with the relaxed fractional atomic coordinates is almost identical with the total density of states computed for the experimental structure. This is not the case for the same calculations in non-spin polarized DFT, see Figs. 3(c) and 3(d). Thus, based on these calculations we can conclude that the coupling between the electronic and lattice degrees of freedom is captured much better in eDMFT than in DFT.

Previously published successful results of structural relaxation in a paramagnetic metallic state, in an insulating ordered state [45], together with current structural relaxations in the paramagnetic insulating state, strengthen further the predictive power of the eDMFT method for the electronic and structural properties at all temperatures in correlated materials.

### C. LT structural changes

As we discussed previously, our room temperature IR and Raman phonon data agree well with the group theory predictions based on the $P6_3mc$ space group which was found by the single crystal X-ray diffraction measurements. As far as we know, there are no reports about the symmetry at low temperature. From the measurements of the electrical polarization, see Fig. 1(d), we observed that concomitant with the magnetic ordering there is an increase in the electrical polarization. This increase suggests that there are structural changes at the magnetic transition in both compounds. Since these materials are pyroelectric, which means that the change in electrical polarization can happen without breaking the symmetry, it is not clear whether the structural changes at the magnetic transition are due to (I) an isostructural phase transition (no change of symmetry away from $P6_3mc$) or (II) a true structural phase transition (the symmetry is lowered from $P6_3mc$). In order to shed light on this matter we have measured low temperature phonon spectra and compared them with the high temperature data.

As temperature lowers below $T_N(Fe) = 60$ K, three new IR-active modes ($E_1$ modes at 270, 429 and 470 cm$^{-1}$) and eight new Raman-active modes ($A_1$ modes at 232 and 852 cm$^{-1}$, $E_1$ mode at 747 cm$^{-1}$ and $E_2$ modes at 127, 158, 193, 224 and 253 cm$^{-1}$) appear in the spectra of $Fe_2Mo_3O_8$. Thus the number of phonon modes observed at $T<T_N(Fe)$ in the IR spectra is $10A_1 + 12E_1$ and in the Raman spectra is $11A_1 + 12E_1 + 16E$ which is bigger than the predicted number of IR-active ($9A_1 + 12E_1$) and Raman-active ($9A_1 + 12E_1 + 13E_2$) phonons by the group theory for the $P6_3mc$ structure. This indicates that there is a structural phase transition concurrent with the magnetic ordering in $Fe_2Mo_3O_8$ compound which agrees with the conclusion made in Ref. [7]. No spectral changes have been detected for $Mn_2Mo_3O_8$ compound down to 5 K suggesting that there is no symmetry charge at the magnetic ordering. Thus, our data show that while in Fe compound a concomitant structural phase transition occurs at $T_N$, in the Mn compound an isostructural phase transition occurs. The existence of the structural phase transition in Fe compound could be related to the more complex ground state compared to the one in Mn compound or to the presence of spin-orbit coupling which is manifested by the strong Ising-like anisotropy on the magnetic susceptibility data [14].

### D. Temperature dependence of electronic transitions in $Fe^{2+}$ ions in $Fe_2Mo_3O_8$

In Section III.B we argued that the broad feature which is observed in Figs. 9(a) and 9(c) at ~3000 cm$^{-1}$ and redshifts by ~400 cm$^{-1}$ with the temperature increase from 5 to 300 K can be related to *d-d* transitions in Fe$^{2+}$ (o) ions in Fe$_2$Mo$_3$O$_8$. While the line is present in the spectra of Fe$_2$Mo$_3$O$_8$ sample_1 it is not observed in the spectra of Fe$_2$Mo$_3$O$_8$ sample_2. One may also notice that the lines corresponding to *d-d* transitions in Fe$^{2+}$ (t) ions at 3400-3500 cm$^{-1}$ are broader and less intense in the spectra of Fe$_2$Mo$_3$O$_8$ sample_1 than in the spectra of Fe$_2$Mo$_3$O$_8$ sample_2 which presumably indicates that sample_1 has some defects affecting the oxygen environment of Fe$^{2+}$ (t) ions. This may point on a possible link between existence of some distortions in FeO$_4$ tetrahedra and observation of the broad line at ~3000 cm$^{-1}$. The broad line was also observed in the spectra of FeZnMo$_3$O$_8$ where Fe$^{2+}$ (t) ions are substituted with Zn$^{2+}$ ions, which again creates some distortion in the tetrahedra due to different ionic radii of Fe$^{2+}$ and Zn$^{2+}$ ions. Since FeO$_6$ octahedra and FeO$_4$ tetrahedra are connected through a common oxygen ion, distortions of the tetrahedra will affect the crystal field at the octahedral sites. We may argue that observation of a broad feature at ~3000 cm$^{-1}$ is related to some phase which is competing with another phase and switching between the two can be triggered by distortions introduced by either defects or Zn doping. Gigantic value of energy shift (~400 cm$^{-1}$) of the broad feature with temperature and strong temperature dependence of its intensity suggest that the phase in which the feature is observed can be also controlled with temperature. In one of the possible scenarios we may imagine two electronic states with close energy levels competing for the ground state so that the latter can be tuned by either defects or Zn doping or temperature. Further theoretical and experimental studies are required to shed more light on the origin of the observed spectral feature at ~3000 cm$^{-1}$.

The group of narrow lines which is observed in 3400-3500 cm$^{-1}$ region for Fe$_2$Mo$_3$O$_8$ [see Figs. 6(a)-6(b)] we attribute to *d-d* electronic transitions in tetrahedrally coordinated Fe$^{2+}$ ions. To better understand these transitions, we need to consider the energy scheme of Fe$^{2+}$ ions in the crystal-field of the Fe$_2$Mo$_3$O$_8$ matrix. Theoretical treatment of splitting of the energy levels of a free Fe$^{2+}$ ion in a cubic crystal-field was given by Low and Weger in Ref. [46]. The ground state of the free Fe$^{2+}$ (3$d^6$) ion is $^5$D. The crystal-field of the cubic symmetry splits the ground term into orbital doublet $^5E$ and orbital triplet $^5T_2$ states separated by $\Delta=10|Dq|$, where $Dq$ is a constant factor which represents the cubic field strength (see Fig. 13). For the octahedrally coordinated $Fe^{2+}$ ions, the ground state is $^5T_2$ [46], while for the tetrahedral coordination the ground state is $^5E$ [47]. Spin-orbit coupling, which is much weaker than the crystal-field splitting for the $Fe^{2+}$ ions, further splits $^5E$ and $^5T_2$ states resulting in the fine structure of the $Fe^{2+}$ levels.

Let's consider the tetrahedrally coordinated $Fe^{2+}$ ions. Spin-orbit coupling splits the $^5E$ state into 5 equidistant levels separated by $\delta=6\lambda^2/\Delta$ energy gaps, where $\lambda$ is a spin-orbit coupling constant between $^5E$ and $^5T_2$ states [46]. The wave functions of these 5 levels transform according to $\Gamma_1, \Gamma_4, \Gamma_3, \Gamma_5, \Gamma_2$ irreducible representations of the $T_d$ point symmetry group listed in the order of energy. The equidistant structure of the split $^5E$ ground-state was experimentally confirmed for $Fe^{2+}$ ions in the ZnSe, CdTe and $MgAl_2O_4$ compounds with a typical value of the energy gap $\delta$ between 10 and 20 cm$^{-1}$ [47]. The fine-structure of the split $^5T_2$ state is more complex and the levels are no longer equidistant on the energy scale. In the static crystal-field theory approximation, the wave functions of the 6 sub-levels of the $^5T_2$ state transform according to $\Gamma_5, \Gamma_4, \Gamma_3, \Gamma_5, \Gamma_4$ irreducible representations listed in the order of energy increase. The corresponding energy scheme of the tetrahedrally coordinated $Fe^{2+}$ ions is shown in Fig. 13. It was also shown that the $^5T_2$ state is perturbed by the dynamic Jehn-Teller distortion to a much higher degree than the $^5E$ ground-state [47], which can renormalize and even change the order of the split $^5T_2$ levels. While not equidistant, the variation of the energy gap $\delta$ within a split $^5T_2$ state still did not exceed 100% with the typical value of $\delta$ between 10 and 20 cm$^{-1}$ for CdTe and ZnSe and about 65 cm$^{-1}$ for $MgAl_2O_4$ [47]. At low temperatures there is a total of 5 lines observed in the IR and Raman spectra of $Fe_2Mo_3O_8$ in the 3400-3500 cm$^{-1}$ region (see Figs. (6)-(7) and Table II) and they are grouped together with a typical distance between them in the range of 10 to 20 cm$^{-1}$. At $T=5$ K only the ground-state level of the split $^5E$ state should be populated and, thus, we can expect to observe up to 6 spectral lines corresponding to transitions from the ground level to the split levels of $^5T_2$ state. We believe that the 5 spectral lines observed in IR and Raman spectra in 3440–3500 cm$^{-1}$ region at $T=5$ K correspond to the above-mentioned transitions from the ground level of the split $^5E$ state. We note that 3440 cm$^{-1}$ line is separated from the 3448 cm$^{-1}$ line by only 8 cm$^{-1}$ and thus can be also related to the transition from the first excited level rather than from the ground level of the split $^5E$ state. In this case the first excited level of the $^5E$ state would be at 8 cm$^{-1}$ which can still be appreciably populated at 5 K resulting in 3440 cm$^{-1}$ transition. To clarify the origin of the 3440 cm$^{-1}$ transition further spectral measurements below 5 K would be beneficial. As temperature increases, the excited levels of $^5E$ state become populated, which results in transitions from these levels to the levels of the $^5T_2$ state. These transitions are manifested as lower frequency satellites of the spectral lines corresponding to transitions from the ground state. We tentatively assign the spectral lines which appear with temperature increase at 3358 and 3405 cm$^{-1}$ (see inset of Fig. 11) to transitions from the excited levels of the $^5E$ state to the levels of the $^5T_2$ state. These lines are redshifted from the nearest to them 3440 cm$^{-1}$ line observed in the Raman spectra at 5 K [see Fig. 10(d)] by 82 and 35 cm$^{-1}$ correspondingly, indicating that the $^5E$ state should be split by at least 82 cm$^{-1}$ and may possess energy levels at 35 and 82 cm$^{-1}$.

Vanishing of infrared and Raman modes corresponding to electronic $d$-$d$ transitions in Fe$^{2+}$ (t) ions above the temperature of structural and magnetic ordering phase transitions $T_N$(Fe) = 60 K may be related to the changes in the symmetry of the Fe$^{2+}$ (t) environment induced by the phase transitions.

## V. CONCLUSIONS

Optical properties and lattice dynamics of hexagonal $M_2Mo_3O_8$ ($M$ = Fe, Mn) single crystals have been studied experimentally in a wide temperature range by means of infrared reflectivity and Raman scattering. At 85 K, i.e. above the magnetic ordering temperature for both compounds, the far-IR spectra of Fe (Mn) compound reveal 19 (22) IR-active phonons, 10 (9) of them are polarized along the $c$-axis and 9 (13) are polarized within the $a-b$ plane. Raman measurements revealed 9(10), 11(12) and 11(12) Raman-active phonons in $a(c,c)\bar{a}$, $a(c,b)\bar{a}$ and $c(a,b)\bar{c}$ configurations correspondingly for Fe$_2$Mo$_3$O$_8$ (Mn$_2$Mo$_3$O$_8$) compound. Group theoretical mode analysis and complimentary density functional theory lattice dynamics calculations are consistent with the $M_2Mo_3O_8$ structure belonging to the high temperature $P6_3mc$ space group determined from single crystal X-ray scattering. All observed vibrational modes are assigned to the specific eigenmodes of the lattice. Electronic and structural properties are well reproduced within eDMFT method for the paramagnetic insulator. These results together with previously published results for other electronic states (such as paramagnetic metal and magnetic insulator), proves the predictive power of the eDMFT method in correlated materials, at finite temperatures, over a large electronic, magnetic and structural phase space.

At temperatures below $T_N$(Fe) = 60 K, several additional IR- and Raman-active modes are detected in experimental spectra of $Fe_2Mo_3O_8$ compound. This confirms the occurrence of a structural transition in $Fe_2Mo_3O_8$ crystal concurrent with the magnetic ordering of Fe spins. On the other hand no spectral changes have been detected for $Mn_2Mo_3O_8$ compound down to 5 K while the changes in polarization $\Delta P(T)$ are similar for both Mn and Fe compounds as they cross the magnetic phase transition. This suggests that in Mn compound magnetic ordering is concurrent with an isostructural phase transition. Unusual broad spectral feature has been observed in infrared spectra of $Fe_2Mo_3O_8$ crystal at ~3000 cm$^{-1}$ and it demonstrates a large frequency shift by ~400 cm$^{-1}$ as the temperature changes in 5-300 K range. We attribute it to *d-d* electronic transitions in $Fe^{2+}$ ions in octahedral coordination. We have also found 5 narrow modes in the IR and Raman spectra in 3400 – 3500 cm$^{-1}$ range in $Fe_2Mo_3O_8$ compound at $T$=5 K. The modes are separated by a temperature-independent distance of 10-20 cm$^{-1}$ from each other. We attribute them to optical transitions from the ground-state to the split levels of $^5T_2$ state of the single $Fe^{2+}$ ions in tetrahedral coordination.

## ACKNOWLEDGMENTS

The Raman scattering and IR reflectivity measurements and sample growth by T.N.S., A.A.S., Y.W. and S.W.C. and V.K. were supported by the U.S. Department Energy DOE DE-FG02-07ER46382. Z.L., NSLS-II, BNL acknowledges DE-AC98-06CH10886 and CDAC DE-NA-0002006 support. The State of Texas via the Texas Center for Superconductivity provided support to A.P.L. for theoretical calculations. Near-IR-VIS ellipsometry measurements at the Center for Functional Nanomaterials, Brookhaven National

Laboratory, have been supported by DOE DE-AC02-98CH10886. The DFT and eDMFT work was supported by the U.S. Department of Energy, Office of Science, Basic Energy Sciences, as a part of the Computational Materials Science Program, funded by the U.S. Department of Energy, Office of Science, Basic Energy Sciences, Materials Sciences and Engineering Division in the case of G.L.P. and by NSF-DMR 1709229 in the case of K.H. Access to the x-ray facilities at the Research Complex, the Rutherford Appleton Laboratory is gratefully acknowledged. The authors are thankful to K. H. Ahn and E. Nowadnick and T.A. Tyson at NJIT for useful discussions.

TABLE I. Internal structural parameters obtained experimentally at room temperature and by using various relaxations methods for $Mn_2Mo_3O_8$ panel (a) and $Fe_2Mo_3O_8$ panel (b). During the structural relaxations the lattice parameters were kept fixed to those reported in the Table.

| (a) | $Mn_2Mo_3O_8$ $P6_3mc$ No.186 $a = b = 5.79750$ Å $c = 10.27070$ Å $\alpha = \beta = 90°$ $\gamma = 120°$ | | | | | | | | |
|---|---|---|---|---|---|---|---|---|---|
| | Experiment | | | eDMFT | | | DFT | | |
| | X | Y | Z | X | Y | Z | X | Y | Z |
| Mn1 | 1/3 | 2/3 | 0.0659 | 1/3 | 2/3 | 0.0651 | 1/3 | 2/3 | 0.0642 |
| Mn2 | 1/3 | 2/3 | 0.6277 | 1/3 | 2/3 | 0.6220 | 1/3 | 2/3 | 0.6405 |
| Mo | 0.1459 | 0.8541 | 0.3647 | 0.1455 | 0.8544 | 0.3614 | 0.1467 | 0.8532 | 0.3646 |
| O1 | 0 | 0 | 0.0000 | 0 | 0 | 0.0000 | 0 | 0 | 0.0000 |
| O2 | 1/3 | 2/3 | 0.2649 | 1/3 | 2/3 | 0.2619 | 1/3 | 2/3 | 0.2536 |
| O3 | 0.4878 | 0.5122 | 0.4719 | 0.4875 | 0.5125 | 0.4708 | 0.4901 | 0.5099 | 0.4978 |
| O4 | 0.8356 | 0.1644 | 0.2492 | 0.8358 | 0.1641 | 0.2458 | 0.8212 | 0.1786 | 0.2521 |

| (b) | $Fe_2Mo_3O_8$ $P6_3mc$ No.186 $a = b = 5.77530$ Å $c = 10.05880$ Å $\alpha = \beta = 90°$ $\gamma = 120°$ | | | | | | | | |
|---|---|---|---|---|---|---|---|---|---|
| | Experiment | | | eDMFT | | | DFT | | |
| | X | Y | Z | X | Y | Z | X | Y | Z |
| Fe1 | 1/3 | 2/3 | 0.0623 | 1/3 | 2/3 | 0.0625 | 1/3 | 2/3 | 0.0611 |
| Fe2 | 1/3 | 2/3 | 0.6226 | 1/3 | 2/3 | 0.6233 | 1/3 | 2/3 | 0.6314 |
| Mo | 0.1460 | 0.8539 | 0.3603 | 0.1457 | 0.8543 | 0.3616 | 0.1474 | 0.8526 | 0.3617 |
| O1 | 0 | 0 | 0.0000 | 0 | 0 | 0.0000 | 0 | 0 | 0.0000 |
| O2 | 1/3 | 2/3 | 0.2584 | 1/3 | 2/3 | 0.2600 | 1/3 | 2/3 | 0.2552 |
| O3 | 0.4869 | 0.5130 | 0.4722 | 0.4871 | 0.5190 | 0.4743 | 0.4881 | 0.5119 | 0.4931 |
| O4 | 0.8337 | 0.1662 | 0.2463 | 0.8333 | 0.1667 | 0.2466 | 0.8228 | 0.1772 | 0.2483 |

TABLE II. Frequencies and symmetries of the experimental IR- and Raman-active phonons at 85 and 5 K, i.e. above and below the magnetic ordering temperature of $M$ spins, as well as calculated (Calc.) phonons in $M_2Mo_3O_8$ ($M$ = Fe, Mn). All phonon frequencies are in cm$^{-1}$. $\varepsilon_{\infty,ab(c)}$ is the value of the dielectric function along the $ab$ plane ($c$-axis) at frequencies higher than that of the highest-energy phonon mode and is listed at the bottom. Modes which appear in the spectra only below the magnetic ordering temperature of $M$ spins are marked with *.

| $M$ | | $A_1$ | | | | | $E_1$ | | | | | $E_2$ | | |
|---|---|---|---|---|---|---|---|---|---|---|---|---|---|---|
| | | IR | | Raman | | Calc. | IR | | Raman | | Calc. | Raman | | Calc. |
| | | 85K | 5K | 85K | 5K | 300K | 85K | 5K | 85K | 5K | 300K | 85K | 5K | 300K |
| Fe | 1 | 230 | 230 | | 232* | 232 | 139 | 129 | 169 | 179 | | | 127* | |
| | 2 | 270 | 269 | 260 | 263 | 267 | | | 191 | 194 | | | 158* | |
| | 3 | 371 | 370 | 368 | 369 | 362 | 218 | 214 | 216 | 213 | 211 | 176 | 180 | 147 |
| | 4 | 447 | 447 | 445 | 446 | 446 | | | 242 | 242 | 264 | | 193* | |
| | 5 | 454 | 453 | 453 | 453 | 514 | | 270* | 264 | 253 | 268 | 211 | 205 | 204 |
| | 6 | 556 | 556 | 553 | 553 | 620 | 291 | 292 | | | 281 | | 224* | |
| | 7 | 646 | 646 | 643 | 644 | | 336 | 336 | 327 | 333 | 338 | 240 | 253* | 240 |
| | 8 | | | 668 | 668 | | | 429* | | | | 267 | 268 | 278 |
| | 9 | 725 | 725 | 724 | 724 | 728 | 452 | 452 | 451 | 454 | 451 | 328 | 328 | |
| | 10 | 770 | 770 | 769 | 771 | 792 | | 470* | | | | 333 | 334 | 367 |
| | 11 | 853 | 853 | | 852* | 838 | 473 | 472 | 481 | 487 | 470 | 448 | 448 | 433 |
| | 12 | | | | | | 514 | 515 | 504 | 500 | 506 | 466 | 469 | 464 |
| | | | | | | | | | | | 549 | | | |
| | 13 | | | | | | 569 | 558 | 565 | 575 | 559 | 513 | 513 | 508 |
| | | | | | | | | | | | 588 | | | |
| | | | | | | | | | | | 660 | | | |
| | 14 | | | | | | | | | 747* | | 555 | 555 | 548 |
| | | | | | | | | | | | | | | 560 |
| | | | | | | | | | | | | | | 572 |
| | | | | | | | | | | | | | | 665 |
| | 15 | | | | | | 743 | 750 | 769 | 769 | 803 | 737 | 746 | 802 |
| Mn | 1 | 204 | 204 | | | 240 | 161 | 161 | 159 | 159 | | 140 | 140 | 141 |
| | 2 | 261 | 262 | 252 | 252 | 296 | 188 | 188 | 187 | 186 | | 184 | 183 | 214 |
| | 3 | 371 | 370 | 369 | 369 | 344 | 221 | 221 | 219 | 219 | 210 | 214 | 214 | 230 |
| | | | | | | 366 | | | | | | | | |
| | 4 | | | 442 | 442 | | 274 | 274 | 272 | 272 | 248 | 265 | 265 | 264 |
| | | | | | | | | | | | 271 | | | |
| | 5 | 456 | 456 | 454 | 454 | 454 | 307 | 308 | 306 | 307 | 301 | 323 | 324 | 308 |
| | 6 | 546 | 547 | 544 | 544 | 504 | 340 | 341 | 339 | 339 | 397 | 343 | 343 | 379 |
| | 7 | 639 | 640 | 637 | 637 | 560 | 443 | 443 | 442 | 442 | 410 | 442 | 442 | 423 |
| | | | | | | | | | | | | | | 430 |
| | 8 | | | 668 | 668 | | 460 | 460 | 459 | 459 | 434 | 461 | 461 | 467 |
| | 9 | 717 | 718 | 717 | 717 | 727 | 475 | 476 | 475 | 475 | 470 | 474 | 474 | 485 |
| | 10 | 780 | 781 | 781 | 781 | 759 | 490 | 490 | | | 483 | 514 | 514 | 518 |
| | 11 | 844 | 842 | 841 | 841 | | 514 | 513 | 513 | 513 | 523 | 555 | 555 | 543 |
| | 12 | | | | | | 562 | 561 | 560 | 560 | 539 | 733 | 733 | 694 |
| | 13 | | | | | | 729 | 729 | | | 690 | | | |
| | 14 | | | | | | | | 781 | 781 | | | | |

Fe:     $\varepsilon_{\infty,ab} = 4.7$, $\varepsilon_{\infty,c} = 5.0$
Mn:    $\varepsilon_{\infty,ab} = 6.1$, $\varepsilon_{\infty,c} = 4.0$

TABLE III. Frequencies $E$ (cm$^{-1}$) of the IR- and Raman-active $d$-$d$ electronic transitions in Fe$^{2+}$ ions in tetrahedral (t) coordination in Fe$_2$Mo$_3$O$_8$ at 5 K along with optical configurations in which they were observed. The notation of the transitions is explained in Fig. 13.

|  |  | IR |  | Raman |  |  |  |
| --- | --- | --- | --- | --- | --- | --- | --- |
| Transition | $E$ (cm$^{-1}$) | $e \perp c$ | $e \parallel c$ | $a(c,c)\bar{a}$ | $a(c,b)\bar{a}$ | $c(a,b)\bar{c}$ | $c(a,a)\bar{c}$ |
| 1→1' | 3440 |  |  | ● |  |  |  |
| 1→2' | 3448 |  | ● |  |  | ● | ● |
| 1→3' | 3467 | ● |  |  | ● | ● | ● |
| 1→4' | 3481 | ● |  |  | ● | ● | ● |
| 1→5' | 3494 |  |  |  |  | ● | ● |

TABLE IV. Wyckoff position, site symmetry, and irreducible representations of atoms for $M_2$Mo$_3$O$_8$ ($M$ = Fe, Mn) (space group $P6_3mc$)

| Atom | Wyckoff notation | Site symmetry | Irreducible representations |
| --- | --- | --- | --- |
| M1 | 2b | $C_{3v}$ | $A_1 + B_1 + E_1 + E_2$ |
| M2 | 2b | $C_{3v}$ | $A_1 + B_1 + E_1 + E_2$ |
| Mo | 6c | $C_s$ | $2A_1 + A_2 + 2B_1 + B_2 + 3E_1 + 3E_2$ |
| O1 | 2a | $C_{3v}$ | $A_1 + B_1 + E_1 + E_2$ |
| O2 | 2b | $C_{3v}$ | $A_1 + B_1 + E_1 + E_2$ |
| O3 | 6c | $C_s$ | $2A_1 + A_2 + 2B_1 + B_2 + 3E_1 + 3E_2$ |
| O4 | 6c | $C_s$ | $2A_1 + A_2 + 2B_1 + B_2 + 3E_1 + 3E_2$ |

Mode classification
$\Gamma_{\text{acoustic}} = A_1 + E_1$
$\Gamma_{\text{Raman}} = 9A_1(xx, yy, zz) + 12E_1(xz, yz) + 13E_2(xy, xx - yy)$
$\Gamma_{\text{IR}} = 9A_1(z) + 12E_1(x, y)$
$\Gamma_{\text{silent}} = 3A_2 + 10B_1 + 3B_2$

FIGURES

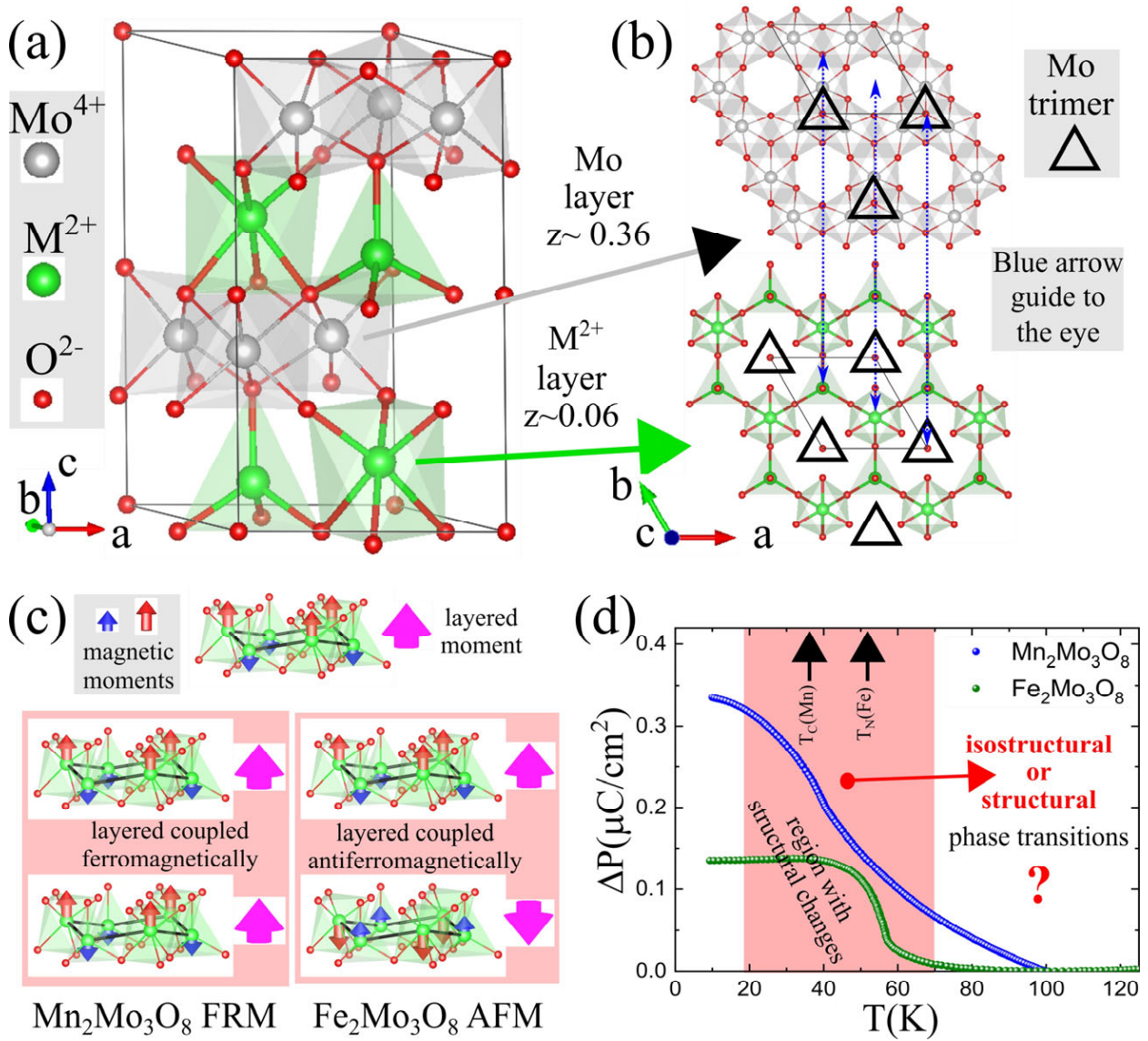

FIG. 1. Schematic overview of the crystal, magnetic and electric properties. (a) Room temperature crystal structure for $M_2Mo_3O_8$; (b) layer arrangements of the $MoO_6$, $MO_6$ and $MO_4$ polyhedra; (c) schematic representation of the magnetic properties; (d) temperature dependence for the variation of the electric polarization for $M_2Mo_3O_8$ ($M$ = Fe, Mn) along the $c$-axis (the data were reproduced from Ref. [9]).

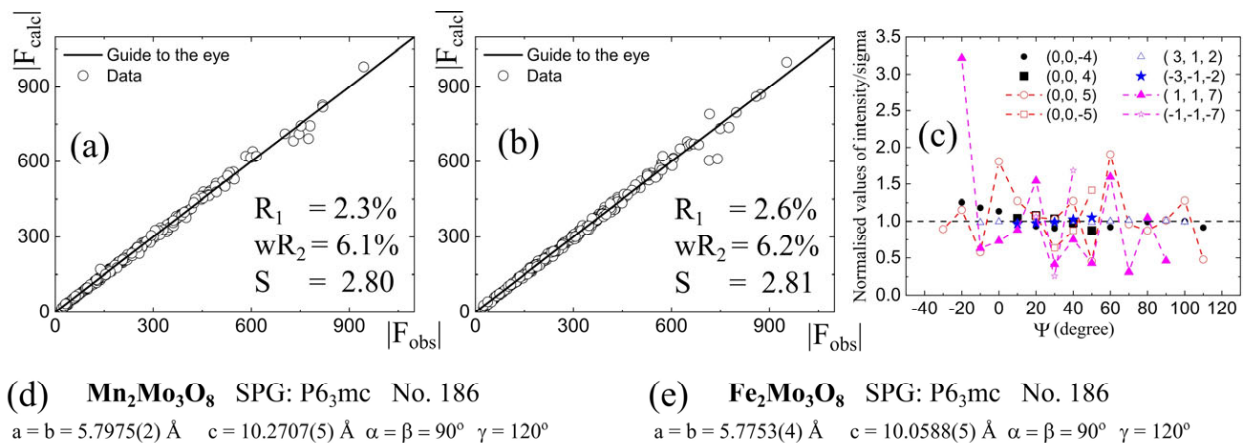

FIG. 2. Structural information from single crystal XRD. Refinement results for (a) $Mn_2Mo_3O_8$ and (b) $Fe_2Mo_3O_8$, comparing calculated and observed structure factors $|F|$; although the refinements were done using $|F|^2$, we plot our results in terms of $|F|$. (c) The azimuthal scan of $Fe_2Mo_3O_8$, showing the normalized values of intensity divided by its sigma. From all the ($h$, $k$, $l$) reflections those with $l$ = even (odd) numbers, indicate allowed (forbidden) reflections. (d)-(e) Refined crystal structure parameters for two compounds with the isotropic thermal parameters assuming the full occupancies.

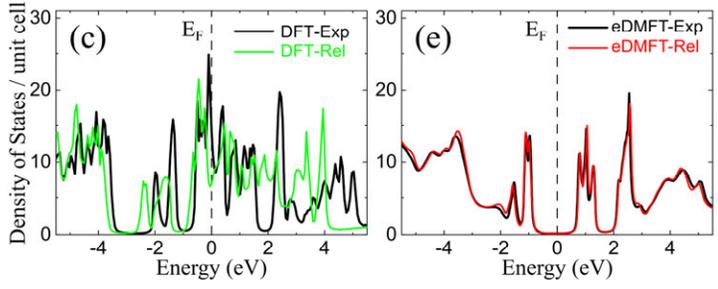
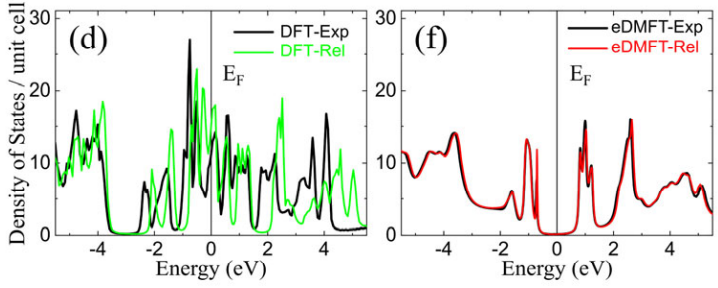
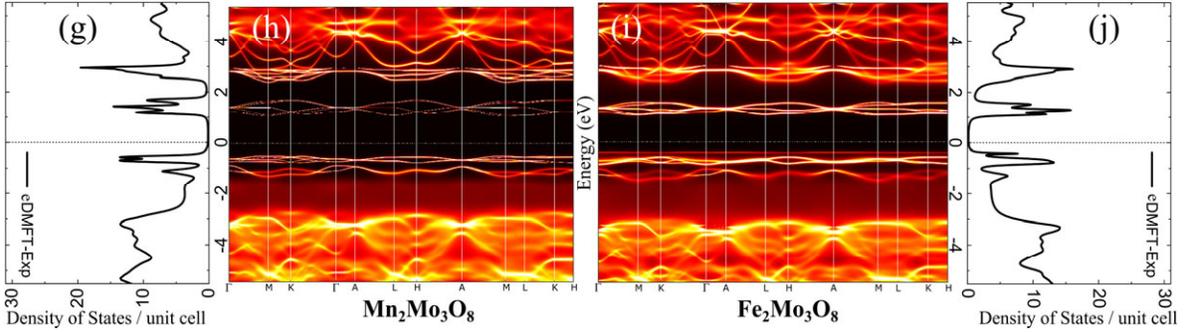

FIG. 3. Structural and electronic properties of $M_2Mo_3O_8$. Panels (a) and (b) show the comparison between the experiment and theoretical methods for various quantities defined in Section III for the transition metal polyhedral ($MO_6$ and $MO_4$); %$_{err}$ represent the percentage error (bold numbers) between the experiment and theory; (c) and (d) show the total density of states obtained by using the DFT method for the experimental structures (black) and for those obtained by using structural relaxations of internal parameters in DFT (green); (e) and (f) show the total density of states obtained by using the DFMT method for the experimental structures (black) and for those obtained by using structural relaxations of internal parameters in DMFT (red); (h) and (i) spectral functions; (g) and (j) repetition of (e) and (f).

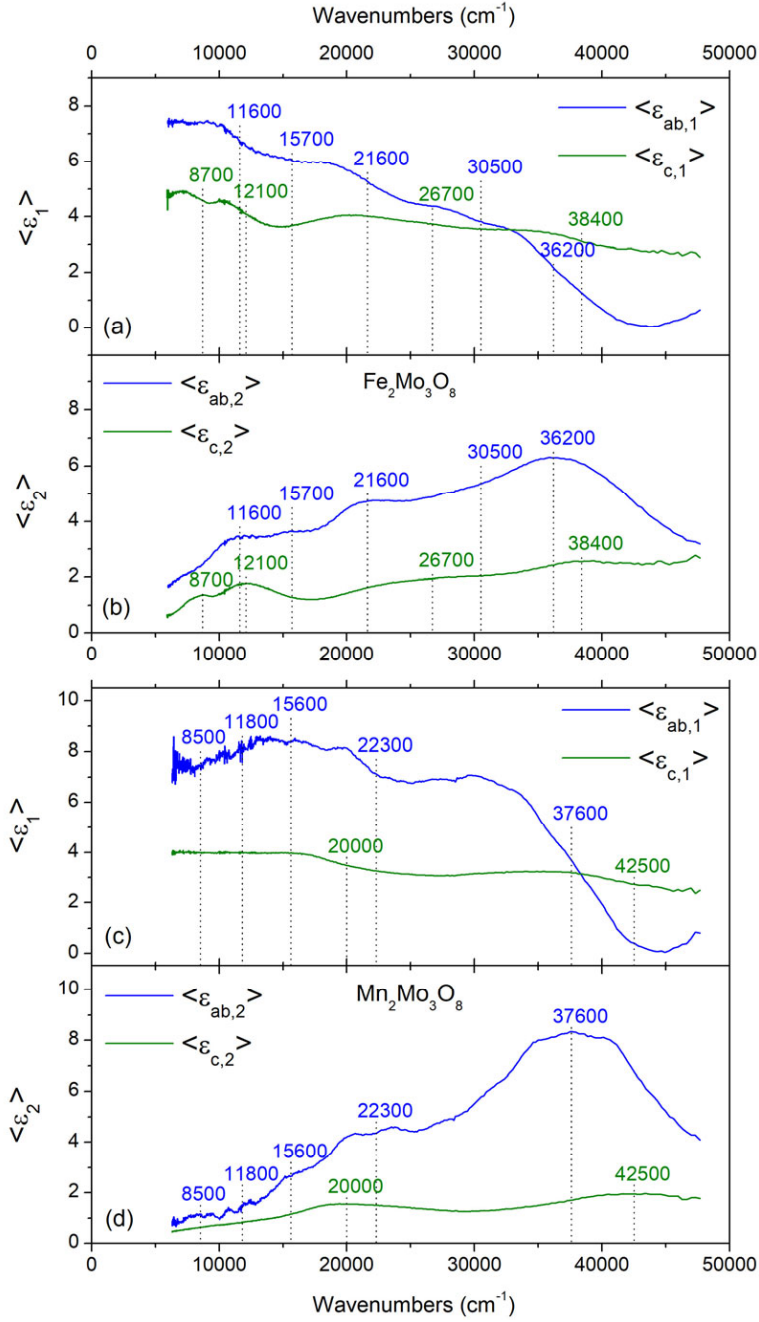

FIG. 4. Spectra of real and imaginary parts of pseudo-dielectric function of (a) $Fe_2Mo_3O_8$ and (b) $Mn_2Mo_3O_8$ in the *ab* plane (blue curve) and along the *c*-axis (olive curve) in the region of electronic *d-d* transitions in $Fe^{2+}$ and $Mo^{4+}$ ions at $T = 300$ K.

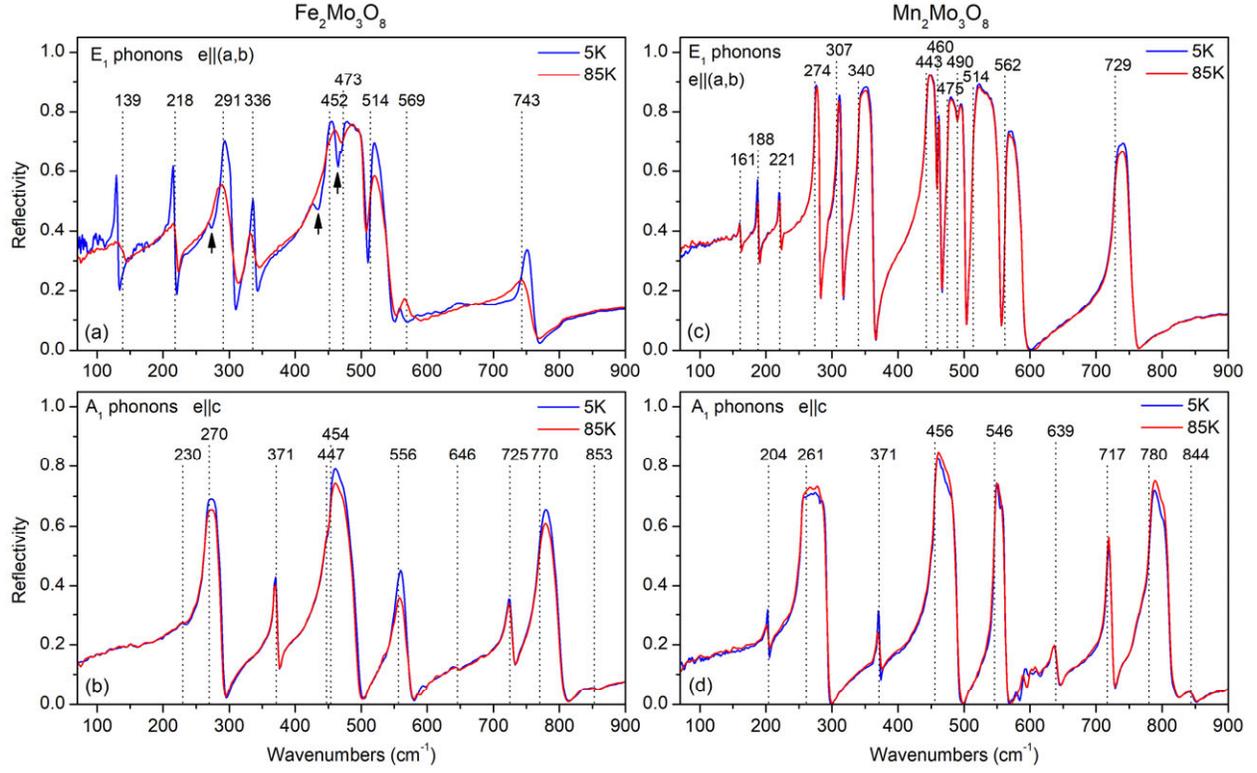

FIG. 5. Reflectivity spectra of (a) and (b) Fe$_2$Mo$_3$O$_8$ and (c) and (d) Mn$_2$Mo$_3$O$_8$ in (a) and (c) $e \perp c$ and (b) and (d) $e \parallel c$ polarizations at 85 K (red line) and 5 K (blue line). Three new lines which appear in the $e \perp c$ spectra of Fe$_2$Mo$_3$O$_8$ below the magnetic ordering temperature of Fe spins $T_N$(Fe) = 60 K are shown with arrows in (a).

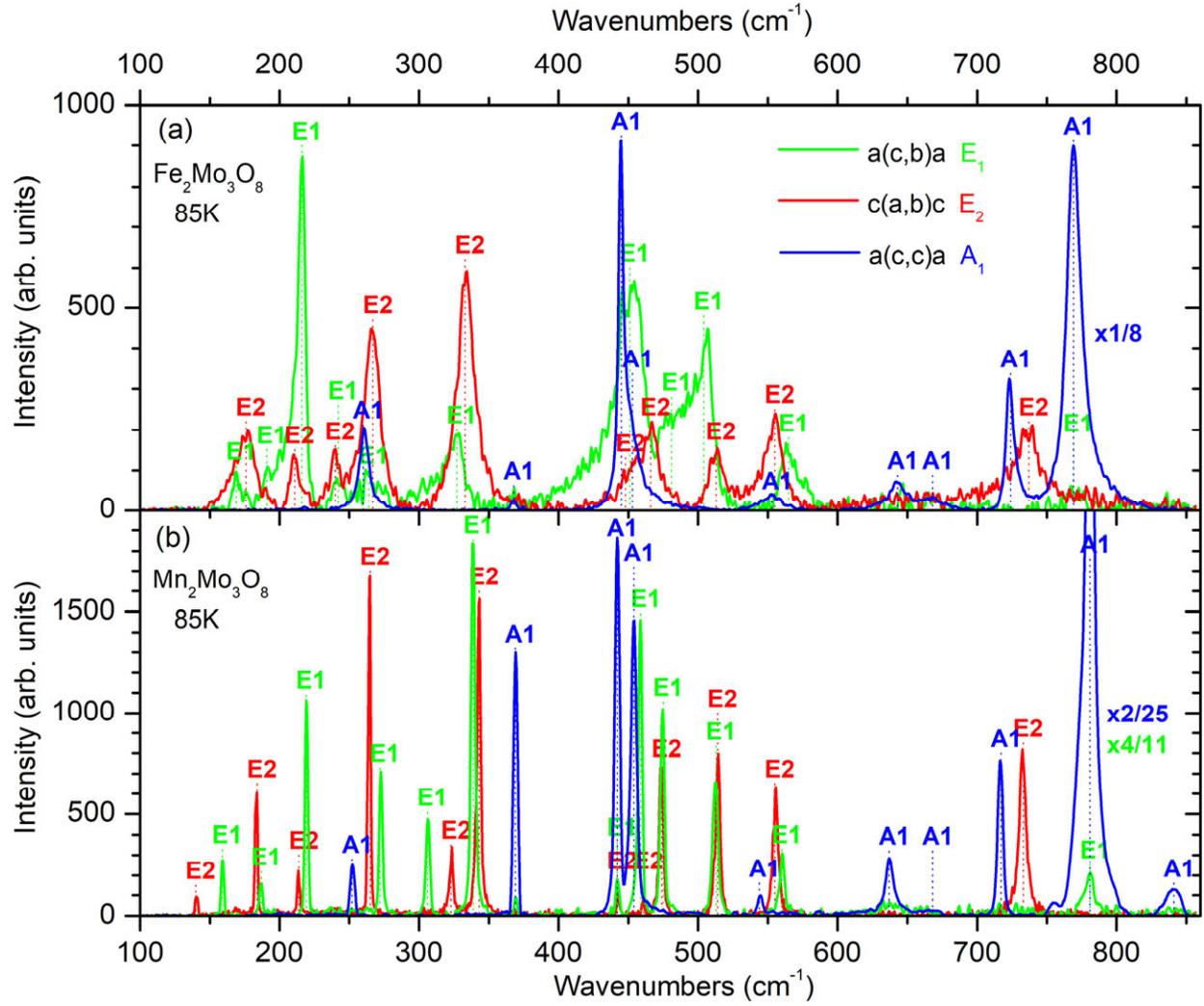

FIG. 6. Raman spectra of (a) $Fe_2Mo_3O_8$ and (b) $Mn_2Mo_3O_8$ measured above the magnetic ordering temperature of Fe (Mn) spins in three complementary scattering geometries: $a(c,b)\bar{a}$ (green lines), $c(a,b)\bar{c}$ (red lines) and $a(c,c)\bar{a}$ (blue lines). The phonon peaks are labelled according to their irreducible representations of $P6_3mc$ space group.

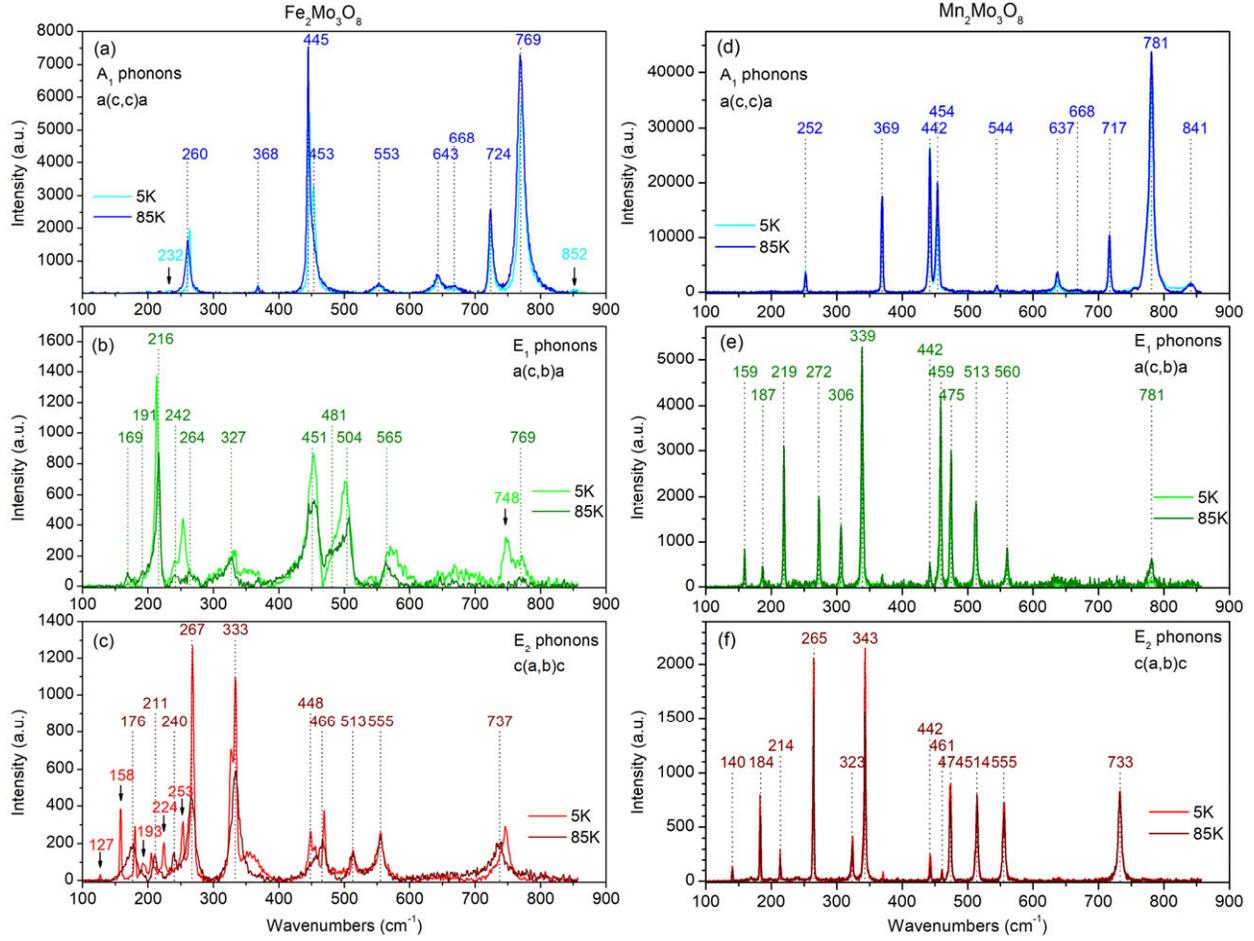

FIG. 7. Raman spectra of phonon modes in (a)-(c) Fe$_2$Mo$_3$O$_8$ and (d)-(f) Mn$_2$Mo$_3$O$_8$ measured in (a) and (d) $a(c,c)\bar{a}$, (b) and (e) $a(c,b)\bar{a}$ and (c) and (f) $c(a,b)\bar{c}$ configurations at 85 and 5 K, i.e. above and below the magnetic ordering temperature of Fe (Mn) spins. New lines which appear in the spectra of Fe$_2$Mo$_3$O$_8$ below the magnetic ordering temperature $T_N$(Fe) = 60 K are marked with arrows in (a)-(c).

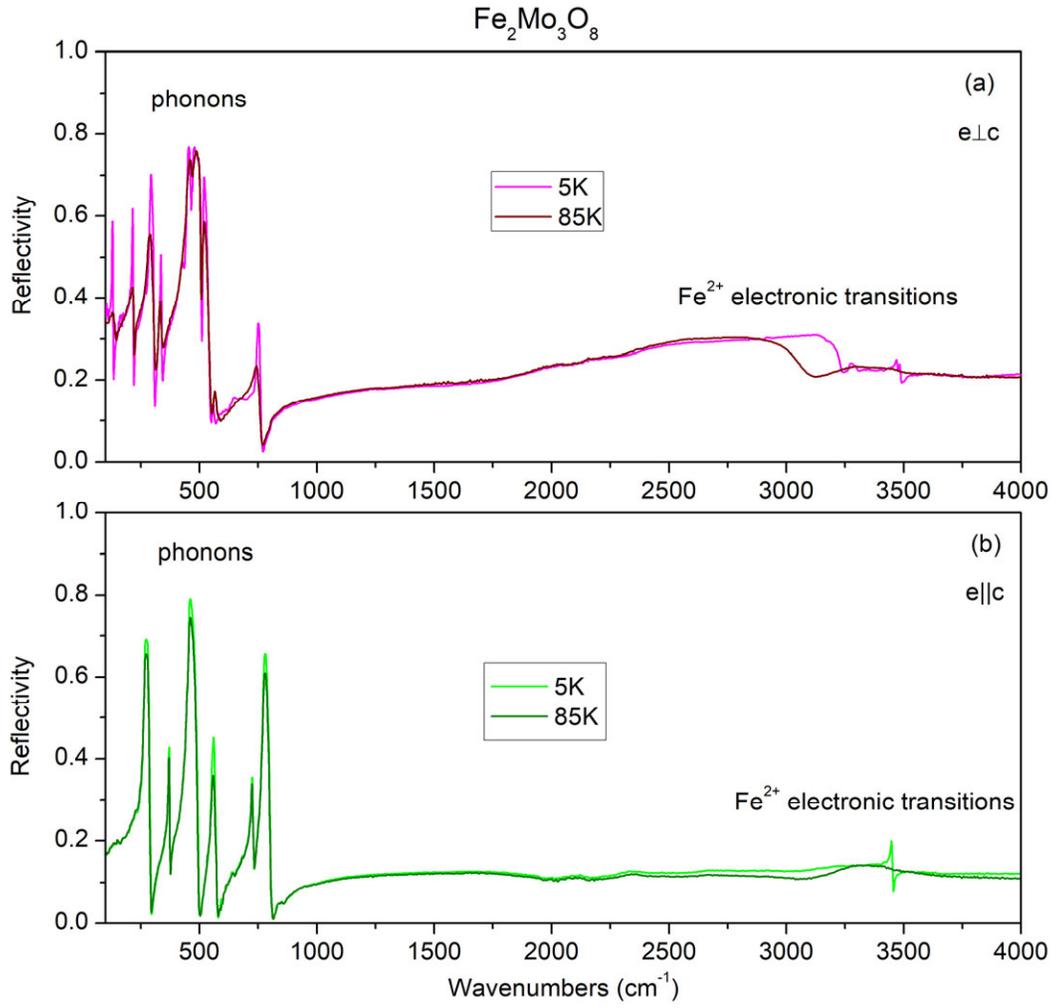

FIG. 8. Infrared reflectivity spectra in the region of phonon modes and electronic $d$-$d$ transitions in $Fe^{2+}$ ions in $Fe_2Mo_3O_8$ crystal in (a) $e \perp c$ and (b) $e || c$ polarizations at 85 and 5 K.

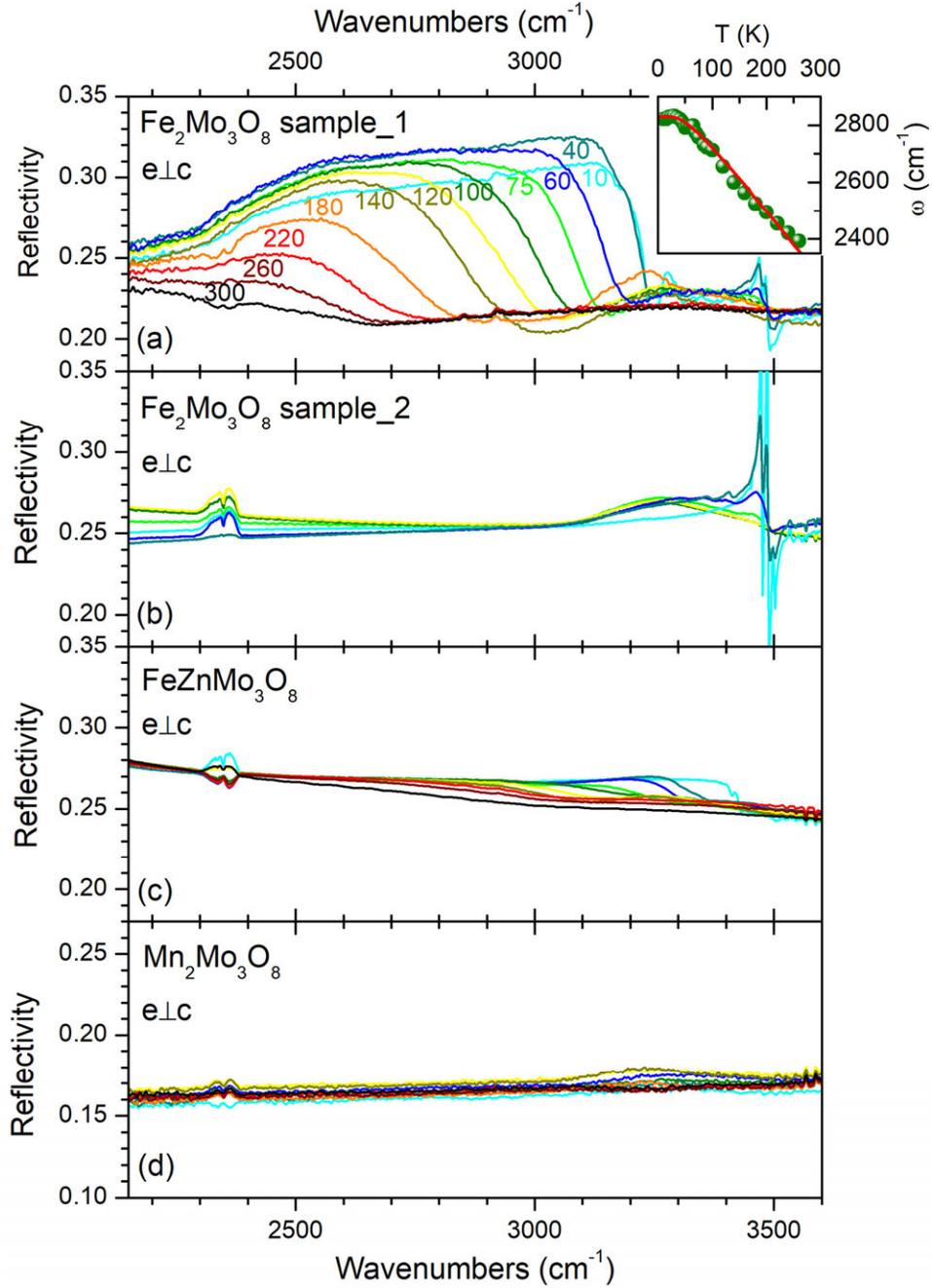

FIG. 9. Infrared reflectivity spectra of (a) $Fe_2Mo_3O_8$ sample_1, (b) $Fe_2Mo_3O_8$ sample_2, (c) $FeZnMo_3O_8$ and (d) $Mn_2Mo_3O_8$ in $e \perp c$ polarization at temperatures between 5 and 300 K. The inset in (a) shows temperature dependence of the position of the broad peak at ~2800 cm$^{-1}$; red line is a guide for the eye.

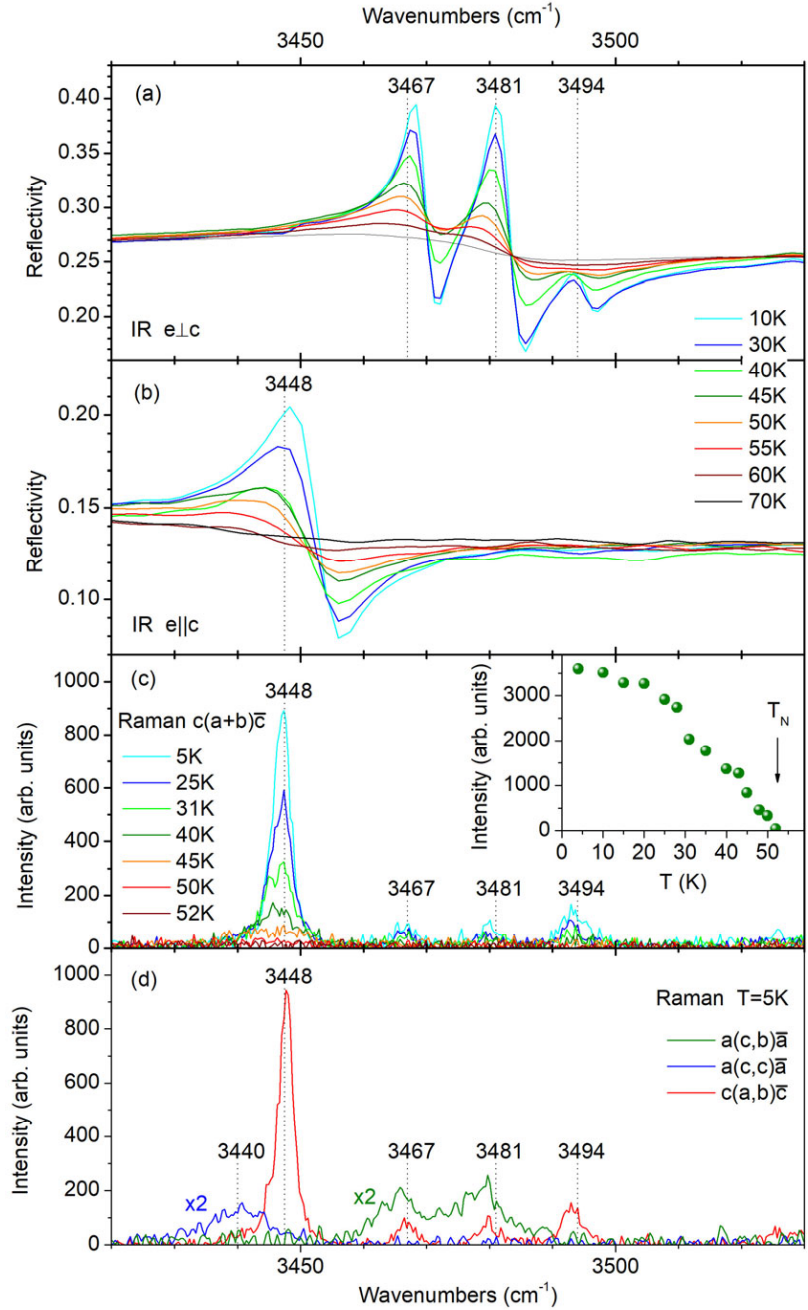

FIG. 10. (a-b) Reflectivity spectra of *d-d* crystal-field transitions in $Fe^{2+}$ ions in $Fe_2Mo_3O_8$ at different temperatures in (a) $e \perp c$ and (b) $e \parallel c$ polarizations. (c-d) Raman spectra of *d-d* crystal-field transitions in $Fe^{2+}$ ions in (c) $c(a+b)\bar{c}$ configuration at different temperatures and (d) $a(c,b)\bar{a}$ (green curve), $a(c,c)\bar{a}$ (blue curve) and $c(a,b)\bar{c}$ (red curve) configurations at 5 K. Inset in (c) shows temperature dependence of the intensity of 3448 cm$^{-1}$ Raman mode.

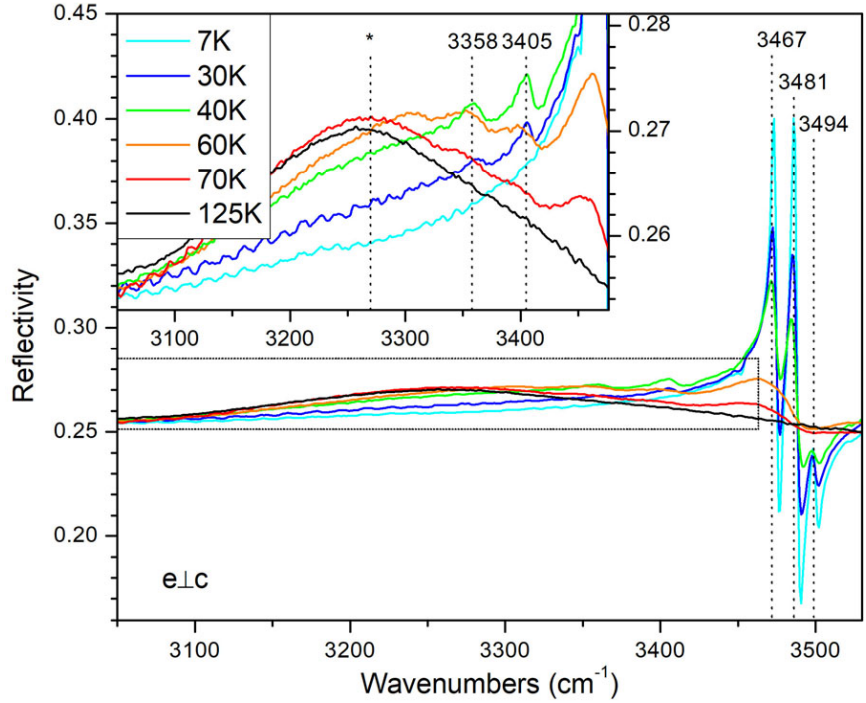

FIG. 11. Reflectivity spectra of *d-d* crystal-field transitions in $Fe^{2+}$ ions in tetrahedral (t) coordination in $Fe_2Mo_3O_8$ at different temperatures in $e{\perp}c$ polarization. Inset is zoom of the rectangular region showing temperature dependent spectra of satellite lines at 3358 and 3405 cm$^{-1}$ appearing with temperature increase at the low energy side of the $Fe^{2+}$ (t) *d-d* transitions observed at 3467, 3481 and 3494 cm$^{-1}$ at *T*=5 K.

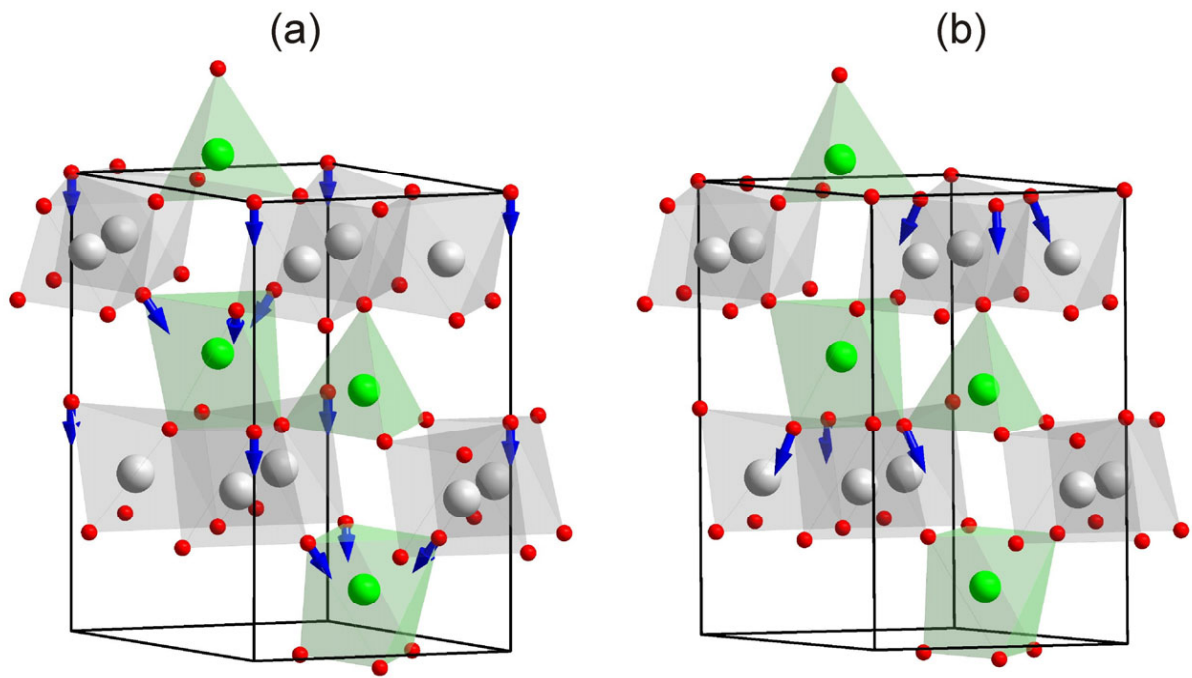

FIG. 12. Atomic displacements corresponding to (a) the highest frequency $A_1$ mode at 852 cm$^{-1}$ and (b) the highest intensity Raman $A_1$ mode at 446 cm$^{-1}$ in Fe$_2$Mo$_3$O$_8$ obtained from the DFT calculations. FeO$_4$ tetrahedra and FeO$_6$ octahedra are shown in green and MoO$_6$ octahedra are shown in grey.

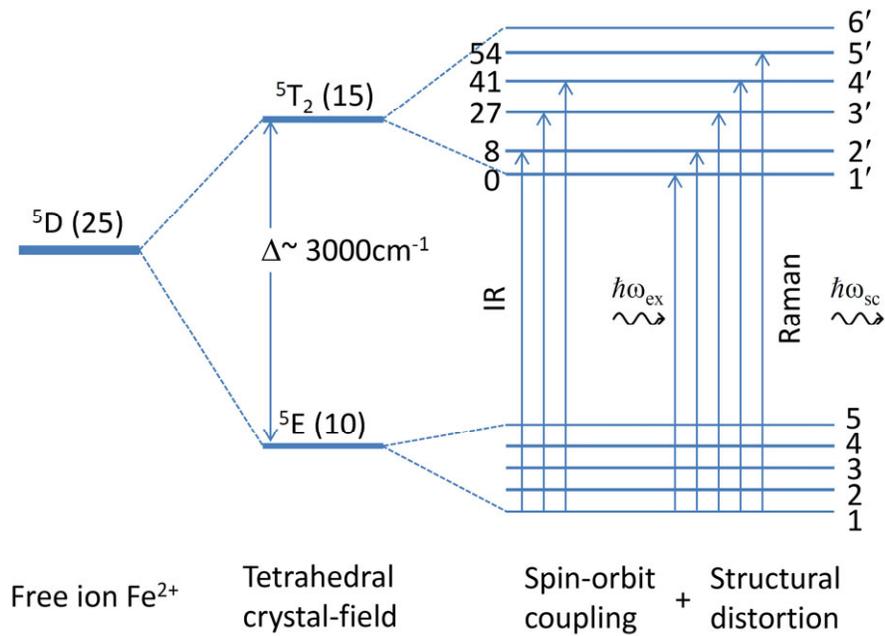

FIG. 13. Energy diagram of the splitting of the ground $^5D$ term of the $Fe^{2+}$ ($3d^6$) free ion by the tetrahedral crystal field and spin-orbit interaction combined with structural distortion below the magnetic ordering temperature of Fe spins $T_N(Fe)=60$ K in $Fe_2Mo_3O_8$. Numbers to the left of the split $^5T_2$ levels indicate their possible relative energies in cm$^{-1}$ as obtained from the analysis of the infrared and Raman spectra.